\documentclass[fleqn,usenatbib]{mnras}

\usepackage{newtxtext,newtxmath}
\usepackage{xcolor}
\usepackage{bm}

\usepackage{color}

\newcommand{\eg}{e.g.,~}
\newcommand{\ie}{i.e.,~}

\usepackage{soul}

\usepackage[T1]{fontenc}

\DeclareRobustCommand{\VAN}[3]{#2}
\let\VANthebibliography\thebibliography
\def\thebibliography{\DeclareRobustCommand{\VAN}[3]{##3}\VANthebibliography}

\usepackage{graphicx}
\usepackage{color}

\usepackage{graphicx}	
\usepackage{amsmath}	
\usepackage[toc,page]{appendix}

\title[Homogeneous turbulence in curved spacetimes]{Measuring the
  properties of homogeneous turbulence in curved spacetimes}

\author[R. Megale et al.]{R. Megale,$^{1}$\thanks{E-mail: rita.megale@unical.it }
A. Cruz-Osorio$^{2}$,
G. Ficarra$^{1}$,
M. Imbrogno$^{1}$,
C. Meringolo$^{3}$,
L. Primavera$^{1}$,
L. Rezzolla$^{3,4,5}$,
\newauthor{and S. Servidio$^{1}$}
\\
$^{1}$ Dipartimento di Fisica, Universit\`a della Calabria,
Arcavacata di Rende (Cosenza), 87036, Italy \\
$^{2}$ Instituto de Astronom\'ia, Universidad Nacional Aut\'onoma de M\'exico, AP 70-264, Ciudad de M\'exico 04510, Mexico \\
$^{3}$ Institut f\"ur Theoretische Physik, Goethe-Universit\"at Frankfurt, Max-von-Laue-Strasse 1, D-60438 Frankfurt am Main, Germany \\
$^{4}$ CERN, Theoretical Physics Department, 1211 Geneva 23, Switzerland \\
$^{5}$ School of Mathematics, Trinity College, Dublin, Ireland
}

\date{Accepted 2025 October 27. Received 2025 October 21; in original form 2025 September 01}

\pubyear{2025}

\begin{document}
\label{firstpage}
\pagerange{\pageref{firstpage}--\pageref{lastpage}}
\maketitle

\begin{abstract}
Turbulence in curved spacetimes in general, and in the vicinity of black
holes (BHs) in particular, represents a poorly understood phenomenon that
is often analysed employing techniques developed for flat spacetimes. We
here propose a novel approach to study turbulence in strong gravitational
fields that is based on the computation of structure functions on generic
manifolds and is thus applicable to arbitrary curved spacetimes. In
particular, we introduce, for the first time, a formalism to compute the
characteristic properties of turbulence, such as the second-order
structure function or the power spectral density, in terms of proper
lengths and volumes and not in terms of coordinate lengths and volumes,
as customarily done. By applying the new approach to the turbulent
rest-mass density field from simulations of magnetised disc accretion
onto a Kerr BH, we inspect in a rigorous way turbulence in regions close
to the event horizon, but also in the disc, the wind, and in the jet. We
demonstrate that the new approach can capture the typical behaviour of an
inertial-range cascade and that differences up to $40-80\%$ emerge in the
vicinity of the event horizon with respect to the standard flat-spacetime
approach. While these differences become smaller at larger distances, our
study highlights that special care needs to be paid when analysing
turbulence in strongly curved spacetimes.
\end{abstract}




\section{Introduction}
\label{sec:introduction}

Accretion discs and the corresponding accumulation of highly dynamical
plasmas in orbital motion are common features of the phenomenology
associated with astrophysical black holes (BHs). Among them, Sagittarius
A* (Sgr A*)~\citep{EHT_SgrA_PaperI_etal, EHT_SgrA_PaperV_etal} and
Messier 87* (M87*)~\citep{Akiyama2019_L1_etal, Akiyama2019_L5_etal} are
the most popular examples of low-luminosity, active galactic nuclei
\citep[see, e.g.,][]{Ho2008, Yuan2014, Algaba2021}. These extraordinary
compact objects produce strong collimated relativistic jets that are very
likely the outcome of mechanisms similar to those proposed by
\citet{Blandford1977} and \citet{Blandford1982}. In such situations, the
strong magnetic fields can extract energy in the form of Poynting flux
from a rotating BH \citep{Takahashi1990}, resulting from the complex
inspiraling plasma motions channelled into the jet region; a rather
similar phenomenology is expected also in ultrarelativistic jets observed
in short gamma-ray bursts~\citep[see, \eg][]{Baiotti2016,
  Paschalidis2016, Murguia-Berthier2016}. Both in laboratory experiments
and in astrophysics, turbulence is expected to play a key role and needs
to be investigated in a suitable theoretical framework, such as
general-relativistic magnetohydrodynamics (GRMHD), widely used among the
fluid approaches.

GRMHD accurately describes the macroscopic plasma dynamics, including
accretion processes near compact objects \citep[see,
  e.g.,][]{Porth2019}. Direct numerical simulations demonstrate that high
Reynolds number turbulence is ubiquitous in accreting BH plasmas, where
fluctuations are often triggered by instabilities and large-scale
inhomogeneities. For example, the magneto-rotational instability is a
crucial mechanism for angular-momentum transport in turbulent accretion
discs, giving rise to chaotic behavior \citep{Balbus1998}. Similarly, the
Kelvin-Helmholtz instability leads to the formation of swirl-like
vortices \citep{Begelman1984}. These instabilities act as
perturbative channels that can initiate magnetic reconnection in the
accretion disc. The latter is a fundamental ingredient of turbulence that
can energise the plasma and accelerate particles to very high energy
\citep{Servidio2009, Servidio2011, Meringolo2023,
  Imbrogno2024}. Given the ubiquity of turbulence in contexts where
spacetime curvature plays an important role and where velocities can
reach relativistic values, it is crucial to be able to analyse the
properties of turbulence and the presence of an active turbulence
cascade.

A fundamental aspect of the theoretical study of turbulence, which is
necessary to understand the distribution of energy among the different
scales involved, is based on the statistics of the fluctuations of
stochastic fields (such as the velocity and the density), as outlined
by~\citet{Frisch1995}. In particular, the second-order structure function
is essential for investigating turbulent environments, as it quantifies
the statistical relationship between differences in physical quantities
at two points separated by a given distance and thus provides a direct
measure of inertial range, both in classical~\citep{Matthaeus2012} and
relativistic fluids~\citep[see, e.g.,~][]{Radice2012b}. Similar
techniques have been widely applied to the solar wind and heliospheric
plasmas, highlighting the presence of the Kolmogorov-like cascade process
\citep{Bruno2005}.

As customary in these studies, the first step involves the computation of
the structure function, and its associated auto-correlation
function. This is followed by the use of the Blackman-Tukey theorem by
means of which it is possible to extract an accurate estimate of the
power spectral density as a function of the wavenumber, by taking the
Fourier transform of the filtered auto-correlation
function~\citep{BlackmanTukey1958}. An important point to remark is that,
although there is a vast literature discussing how to perform such
analyses in classical (i.e., nonrelativistic) turbulence, the exploration
of turbulence in relativistic regimes is much less developed and limited
essentially to flat spacetimes~\citep{Zrake2011a, Radice2012b,
  Zrake2013}.  However, these approaches may be insufficient to
understand the turbulent dynamics of matter under those conditions in
which spacetime curvature plays an important role. Indeed, these are the
typical scenarios of high-energy astrophysics in general and of accretion
discs in particular.

We here analyse turbulence in the vicinity of BHs as computed from
high-resolution two-dimensional numerical simulations performed with
\texttt{BHAC} \citep[]{Porth2017, Olivares2019} of an accretion disc
around a Kerr spacetime. The approach we propose, however, can also be
applied to full three-dimensional (3D) turbulence and be extended to a
general Riemannian manifold. Using these simulations as a reference, we
compute the power spectral density as the distribution of energy over the
wavevectors associated both with coordinate lengths, as done in flat
spacetimes, and with proper lengths. We apply this approach to inspect
the turbulence in four different regions: close to the event horizon, in
the accretion disc, in the wind region, and in the jet. In this way we
can demonstrate that differences up to $40-80\%$ in the properties of the
turbulence can emerge when performing the analysis either in flat or
curved spacetimes. While these differences become smaller at larger
distances, they alert us that special care needs to be paid when
analysing turbulence in strongly curved spacetimes.

The paper is organised as follows. In Sec. \ref{sec:2}, we provide
details about the new technique developed for general relativistic
plasmas. In Sec. \ref{sec:application}, we describe the GRMHD simulation
of a Fishbone-Moncrief torus (FM) \citep{Fishbone1976} set to apply the
technique. In Sec. \ref{sec:pls}, we finally show the results of the
power spectral density associated to the proper measurements of the
turbulence properties. The conclusions are discussed in
Sec.~\ref{sec:conclusions}, while additional validation tests and
complementary analyses are presented in the Appendices. Throughout the
paper, we use geometrised units where $G = c =1$, with $G$ and $c$ being
the gravitational constant and the speed of light, respectively.

\section{Turbulent properties in curved manifolds}
\label{sec:2}

\begin{figure}
  \includegraphics[scale=0.24]{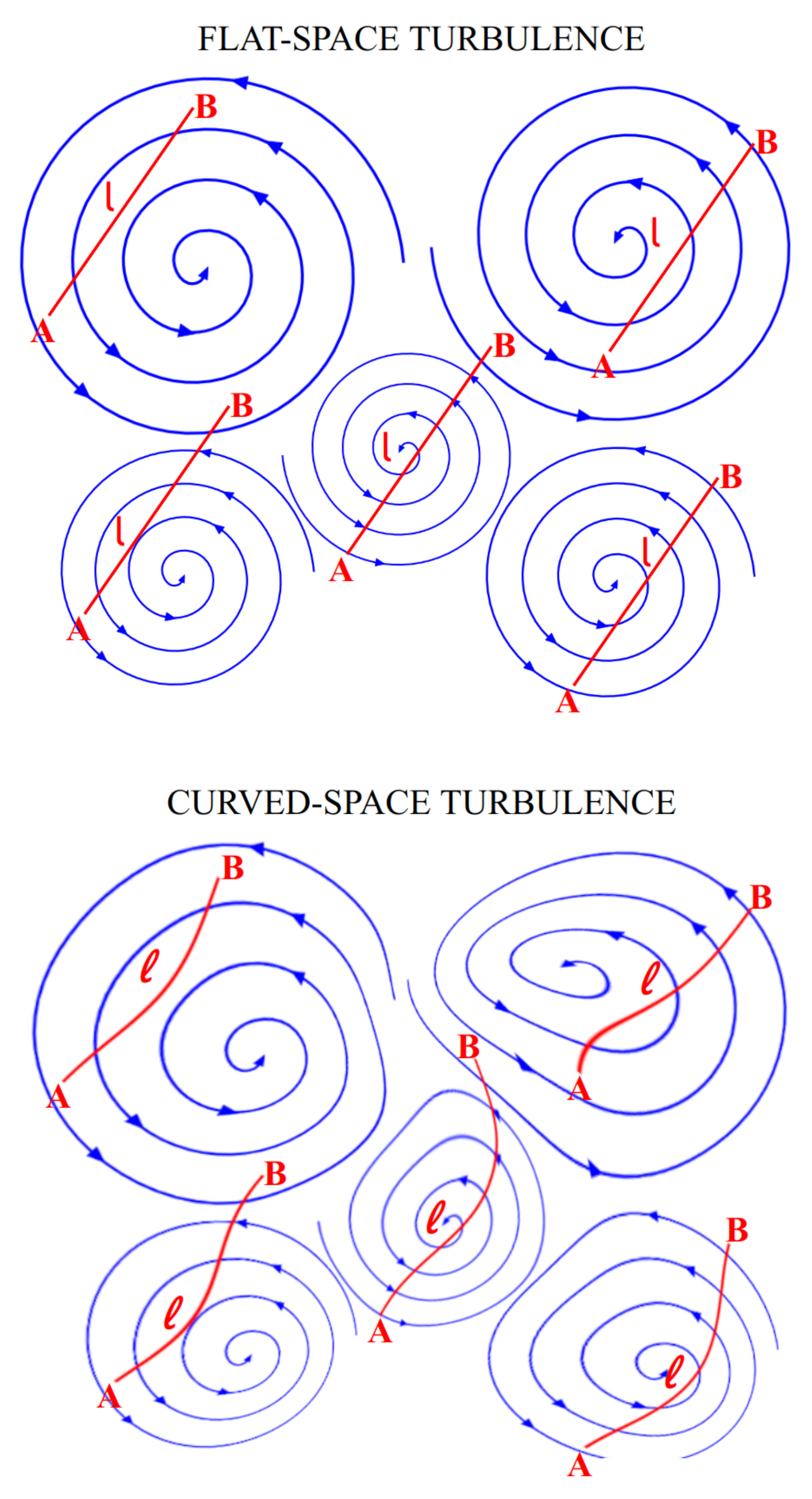}
  \caption{Schematic cartoon contrasting the measurement of turbulent
    structures, \eg vortices indicated in blue, in a flat spacetime (top
    panel) and in a curved one (bottom panel). Note the difference
    between coordinate and proper lengths between two points $A$ and $B$
    in the turbulent field.}
  \label{fig:flat-curved}
\end{figure}

Much of the theory and phenomenology of fully developed and homogeneous
turbulence relies on the ability to correlate physical properties of the
system at different locations, and to measure how these properties, --
e.g., the size of the turbulent eddies -- vary on different scales (see
top panel of Fig.~\ref{fig:flat-curved}). In this respect, a particularly
useful quantity is the so-called ``second-order structure function'',
which measures the differences (or fluctuations) of a given physical
property of the system $\phi$ (e.g., temperature, density, velocity,
etc.) between a representative point $A$ at position $\bm{\vec{x}}_A$
and another point $B$ at separation $\bm{\vec{l}}$ from $A$, i.e., at
$\bm{\vec{x}}_B = \bm{\vec{x}}_A + \bm{\vec{l}}$.

In a stochastic field, as that commonly assumed to characterise classical
homogeneous turbulence, the second-order structure function represents a
robust statistical tool to quantify the level of fluctuations at a given
scale $l$, so that, at any given time, the volume-averaged second-order
structure function can be expressed as~\citep{Frisch1995}
\begin{eqnarray}
  \label{eqn:S2flat}
  S_2(|\bm{\vec{l}}|) &:=& \left \langle \left | f(\bm{\vec{x}}_A +
  \bm{\vec{l}}) - f(\bm{\vec{x}}_A) \right |^2 \right \rangle \nonumber \\
  & = & \left
  \langle \frac{1}{V} \int | f(\bm{\vec{x}}_A + \bm{\vec{l}}) -
  f(\bm{\vec{x}}_A) |^2 d^3 x_A \right \rangle_\Omega\,,
\end{eqnarray}
where the volume integral is performed over the entire spatial volume $V$
that contains homogeneous turbulence and where the brackets $\langle
\quad \rangle_\Omega$ denote an average over the solid angle $\Omega$ of
all the possible paths between $A$ and $B$ separated by the distance
$\bm{\vec{l}}$. It should be noted that in expression~\eqref{eqn:S2flat}
the fluctuations $f$ of the field $\phi$ are evaluated after subtracting
the corresponding volume average, that is,
\begin{equation}
  f := \phi - \frac{1}{V}\int \phi \, d^3 x\,.
\end{equation}

While these concepts and definitions are commonly used and are
well-defined in classical and even special-relativistic frameworks, they
do not take into account the fact that in curved spacetimes lengths and
volumes cannot be expressed in terms of coordinate measurements, but need
to reflect the underlining curved geometry of the manifold (see bottom
panel of Fig.~\ref{fig:flat-curved}). This concept may be best expressed
with a simple example involving again the two representative points $A$
and $B$ in the turbulent field. When considering a standard $3+1$
splitting of spacetime~\citep{ADM2008, Rezzolla2013}, the physically
relevant (i.e., invariant) spatial distance between $A$ and $B$ on a
given spatial hypersurface where they have spatial coordinates
$x^i_{A,B}$ is not given by the coordinate length
\begin{equation}
  l := \int_{x_A^k}^{x_B^k} \sqrt{dx^i dx^j}\,,
  \label{eq:l}
\end{equation}
but rather by the ``proper length'' $\ell$ expressed as
\begin{equation}
    \ell := \int_{x_A^k}^{x_B^k} \sqrt{\gamma_{ij} dx^i dx^j}\,,
     \label{eq:ell}
\end{equation}
where $\gamma_{ij}$ is the spatial three-metric on the constant-time
hypersurface. They are coordinate-independent quantities in the
  sense that a different choice of the spatial metric for the same slice
  would lead to the same proper distances. Obviously,
expressions~\eqref{eq:l} and \eqref{eq:ell} coincide in a flat spacetime
and similar considerations apply also to proper measures of surfaces and
volumes (see below). Bearing this in mind, the general-relativistic (or
curved spacetime) extension of expression~\eqref{eqn:S2flat} will lead to
what we will hereafter refer to as ``proper second-order structure
function''
\begin{equation}
  S_{2,\mathcal{P}}(\ell) := \left \langle \frac{1}{\mathcal{V}} \int
  \left| f(x_B^i) - f(x_A^i) \right|^2 \ \alpha \ \sqrt{\gamma(x_A^i)}
  \ d^3 x_A \right \rangle_{\Omega}\,,
  \label{eqn:S2gamma+geo}
\end{equation}
where $\mathcal{V}$ is the proper volume, i.e., 
\begin{equation}
\mathcal{V} := { \int \alpha \ \sqrt{\gamma} \ d^3 x}\,,
\end{equation}
$\alpha$ is the lapse function (see below for a definition) and $\gamma$
is the determinant of the spatial metric $\gamma_{ij}$, thus accounting
for local variations of the volume in response to the background
curvature. Also in this case, the volume integral in
expression~\eqref{eqn:S2gamma+geo} is computed over the entire spatial
volume that contains homogeneous turbulence, and the average is performed
along all the curves that connect the two points $x_A^i$ and $x_B^i$ and
are separated by a proper length $\ell$. 

Starting from the proper second-order structure function
$S_{2,\mathcal{P}}(\ell)$, it is straightforward to derive a proper
measure of the power spectrum of the fluctuations. In particular, using
the Blackman-Tukey theorem~\citep{BlackmanTukey1958, Frisch1995,
  Matthaeus1982b, Pecora2023}, it is possible to relate the proper
second-order structure function of a turbulent field in a homogeneous
system to the ``proper'' auto-correlation function
$C_{\mathcal{P}}(\ell)$ as
\begin{equation}
  C_{\mathcal{P}}(\ell) := E - \frac{S_{2,\mathcal{P}}(\ell)}{2}\,, 
  \label{eq:cs2}
\end{equation}
where $E$ is a proper-volume average of the energy in the fluctuations of
the field or, equivalently, its variance, i.e.,
\begin{equation}
  E := \frac{1}{\mathcal{V}} \int | f |^2 \ \alpha \ \sqrt{\gamma} \ d^3 x\,.
  \label{eq:variance}
\end{equation}
Note that because $S_{2,\mathcal{P}}(\ell=0)=0$, it follows that
$C_{\mathcal{P}}(\ell=0) = E$, \ie the (proper) auto-correlation function
in the limit of zero-correlation (proper) length is given by the variance
of the energy in the fluctuations.

At this point, it is useful to express the spectral properties of the
proper auto-correlation function~\eqref{eq:cs2}, \ie its power spectral
density (PSD), in terms of the proper length $\ell$ and this can be done
by Fourier-transforming the auto-correlation function after a convolution
with an appropriate windowing function $W_\ell$, i.e.,
\begin{equation}
  \label{eqn:PLS}
    \textrm{PSD}_{\mathcal{P}}(\chi) := \int_{-\infty}^{\infty}
    C_{\mathcal{P}}(\ell) \, W_\ell \, e^{-i\chi \ell} d \ell\,,
\end{equation}
where $\chi := 1/\ell$ and $W_\ell$ is introduced so as to guarantee that
$C_{\mathcal{P}}(\ell)$ has a compact support. More specifically, we have
employed a Hann function, so that $W_\ell := \tfrac{1}{2}\{1-\cos[2\pi
  \ell/(2N_{\text{win}})]\}$, where $N_{\text{win}} = 1024$ is the number
of points used for the window, but our results do not depend sensibly on
this choice.

The method described above is a straightforward extension of
corresponding quantities defined in a flat spacetime and to which they
reduce in the case in which the three-metric $\gamma_{ij}$ is that of a
Minkowski spacetime in a given coordinate system. While what we have
discussed so far is completely generic, in what follows we will apply the
formalism developed so far to the specific case of the turbulence that
appears in GRMHD simulation of disc accretion onto a rotating black hole.

\section{Simulations of accreting BHs}
\label{sec:application}

In order to assess to what extent the use of a flat-spacetime approach
can impact the description of the statistical properties of turbulence in
a curved spacetime, we have considered a well-known scenario on a
turbulent disc accreting onto a rotating black hole and compared the
properties of the second-order structure function when expressed in terms
of a curved or flat spacetime approach. In what follows we briefly review
the mathematical and numerical setup.

\subsection{Mathematical and numerical setup}

As customary in these GRMHD calculations of accretion onto rotating black
holes~\citep[see, \eg][]{Porth2019_etal}, we solve the coupled set of
GRMHD equations in the ideal-MHD limit, i.e.,
\begin{eqnarray}
  &&   \nabla_{\mu} ( \rho u^{\mu})=0\,,\\
  &&   \nabla_{\mu} T^{\mu \nu}=0\,,\\
  &&   \nabla_{\mu} {}^{*}F^{\mu \nu}=0\,,
\end{eqnarray}
where $\rho$ is the rest-mass density, $u^{\mu}$ is the fluid
four-velocity, $T^{\mu \nu}$ is the energy-momentum tensor and
${}^{*}F^{\mu \nu}$ is the dual Faraday tensor~\citep[see, e.g.,][for a
  brief derivation of these equations]{Mizuno2025}.

These equations are solved numerically via the code
\texttt{BHAC}~\citep{Porth2017}, which adopts finite-volume
high-resolution shock-capturing methods to describe the plasma dynamics
in arbitrary but fixed and stationary spacetimes. \texttt{BHAC} employs
adaptive mesh refinement (AMR) techniques using an efficient block-based
approach and a constrained-transport method~\citep{Olivares2019}, which
ensures that the divergence of the magnetic field is maintained to
round-off precision~\citep{DelZanna2007}.

To improve the solution of the GRMHD equations in the vicinity of the
event horizon and avoid the introduction of boundary conditions, the
background spacetime is covered with Kerr-Schild (KS)
coordinates~\citep{Font1998} with the spatial metric in covariant and
contravariant form being given by 
\begin{equation}
\gamma_{ij}=\begin{bmatrix} N & 0 &
-a_* N \sin^2 \theta \\
&&\\
0 & \xi & 0 \\
&&\\
-a_* N \sin^2 \theta & 0 & \sin^2 \theta
\Big[\xi + a_{*}^2 N
  \sin^2 \theta \Big] \\
\end{bmatrix}\,,
\end{equation}
and
\begin{equation}
\gamma^{ij}=\begin{bmatrix} {a_{*}^2 \sin^2 \theta}/{\xi}+
1/N & 0 & {a_*}/{\xi} \\
&&\\
0 & {1}/{\xi} & 0 \\
&&\\
{a_*}/{\xi} & 0 & {1}/({\xi \sin^2 \theta})
\end{bmatrix}\,.
\end{equation}
Here, $a_* := J/M$ is the spin parameter, with $J$ the angular momentum
of the black hole, $M$ the mass, $N:= 1+{2Mr}/{\xi}$, and $\xi := r^2 +
a_{*}^2 \cos^2 \theta$. In these coordinates, the square root of the
determinant of the spatial metric is given by
\begin{equation}
\sqrt{\gamma}=\xi \sqrt{1+ \frac{2Mr}{\xi}} \sin\theta\,,
\end{equation}
while the lapse function and the shift vector, are: $\alpha = \Big( 1 +
2Mr/\xi\Big)^{-1/2}$ and $\beta^i = \Big[ (2Mr/\xi) \Big(1 +
  2Mr/\xi\Big)^{-1}, 0, 0\Big] $ \citep{Alcubierre2008}. To further
increase the resolution near the event horizon, \texttt{BHAC} adopts the
so-called modified Kerr-Schild (MKS) coordinates system \citep{Misner1973,
  McKinney2004}, where two parameters $(s, \lambda)$ are introduced to
stretch the grid radially and near the equatorial region in the polar
direction. The corresponding coordinate transformation is given by
\begin{equation}
  r(s) = R_0 + e^s\,, \quad \theta( \lambda)= \lambda + \frac{h}{2}
  \sin(2\lambda)\,, \quad \tilde{\xi}=e^{2s}+a_{*}^2 \cos^2\theta
\end{equation}
where $R_0$ and $h$ are parameters that control how much resolution is
concentrated near the horizon ($R_0$) and near the equator ($h$). In our
case, $R_0=h=0$ so that the MKS coordinates reduce to the standard
logarithmic KS coordinates and the corresponding
square root of the determinant of the
spatial metric is given by
\begin{equation}
\sqrt{\gamma}= \tilde{\xi} e^s \sqrt{1+
  \frac{2Me^s}{\tilde{\xi}}}\sin\theta\,.
\end{equation}

As initial conditions, and as customary for simulations of this
type~\citep{Porth2019}, we consider an axisymmetric equilibrium
torus with constant specific angular momentum \citep{Fishbone1976}
orbiting a Kerr BH with $a_*=0.9375$. The initial magnetic field is set
to be purely poloidal and specified via the azimuthal vector potential
expressed in terms of the rest-mass density so as the magnetic field is
initially confined in the torus, \ie
\begin{equation}
  A_{\phi} = \text{max} \left ( \frac{\rho}{\rho_{\text{max}}} - 0.99,
    0 \right )\,,
\end{equation}
where $\rho_{\max}=1$. We should note that this prescription corresponds
to a dipolar magnetic with a single neutral line or polarity.  The inner
radius of the torus is chosen to be $r_{\text{in}}=12 \,M$ and the plasma
is assumed to be described by an ideal-gas equation of state with
adiabatic index $\gamma=4/3$~\citep{Rezzolla_book:2013}. To increase the
dynamic range over which the turbulence develops, the simulations are
performed assuming axisymmetry and hence in two spatial dimensions with
five levels of mesh refinement and an effective resolution of $N_r \times
N_{\theta} = 4096 \times 2048$ cells in the radial and polar directions,
respectively. The outer boundary is placed at $5000\,M$ and the evolution
is carried out till time $t=12000\,M$. 

Given our specific combination of parameters, the flow that results is
what is normally referred to as a Standard And Normal Evolution (SANE)
accretion flow and is characterized by a weak magnetic field strength
\citep[e.g.,][]{Narayan2012}. In addition, while the adoption of a
single-polarity magnetic field is very common, it also represents a
matter of choice and different setups are possible where the polarity of
the magnetic field can vary and have alternating multiplicity, which
leads to a rather different global behaviour~\citep[see,
  \eg][]{Nathanail2020}. While we do not expect much of the results
presented here to depend on the initial choice for the magnetic field, it
would be interesting to apply the formalism introduced here also to the
case of magnetic fields having alternating polarity or where the
accretion flow develops following a magnetically-arrested disc (MAD)
phenomenology~\citep{Narayan2000}.

\begin{figure*}
  \includegraphics[width=0.95\textwidth]{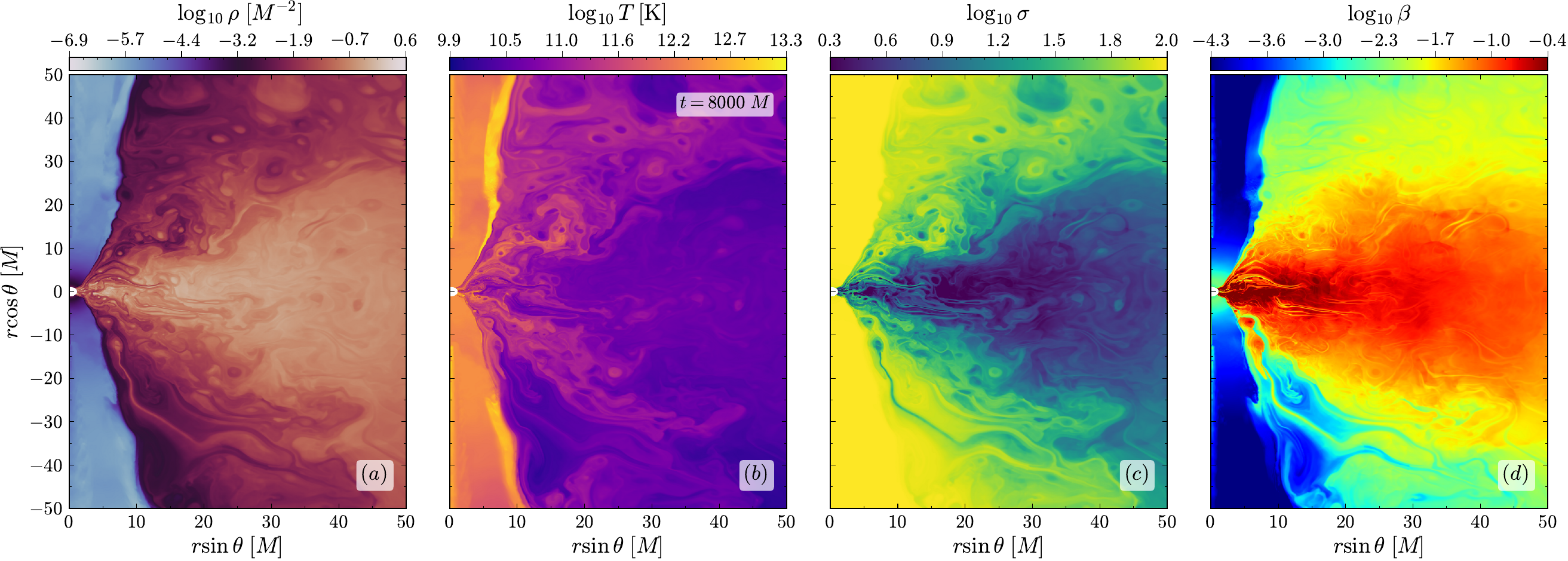}
  \vskip 0.2cm
  \includegraphics[width=0.95\textwidth]{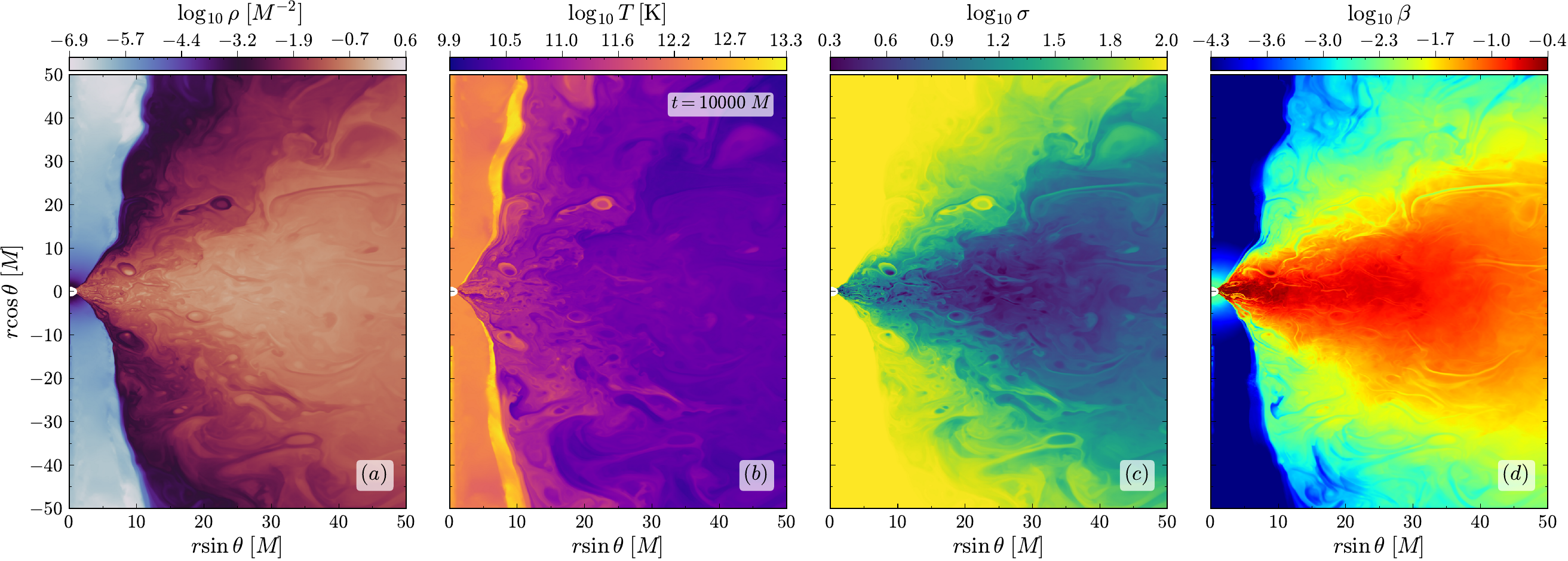}
  \caption{Spatial distributions in a polar slice of some of the most
    relevant plasma quantities at two representative times, $t=8000 \, M$
    (top row) and $t=10000 \, M$ (bottom row). From left to right, are
    reported the rest-mass density $\rho$ [panel $(a)$], the temperature
    $T$ [panel $(b)$], the magnetisation $\sigma$, and the plasma $\beta$
    [panel $(d)$].}
  \label{fig:fields_8000e10000} 
\end{figure*}

\subsection{Turbulent zones near the BH}
\label{subsec:zones}

\begin{figure*}
  \includegraphics[width=0.95\textwidth]{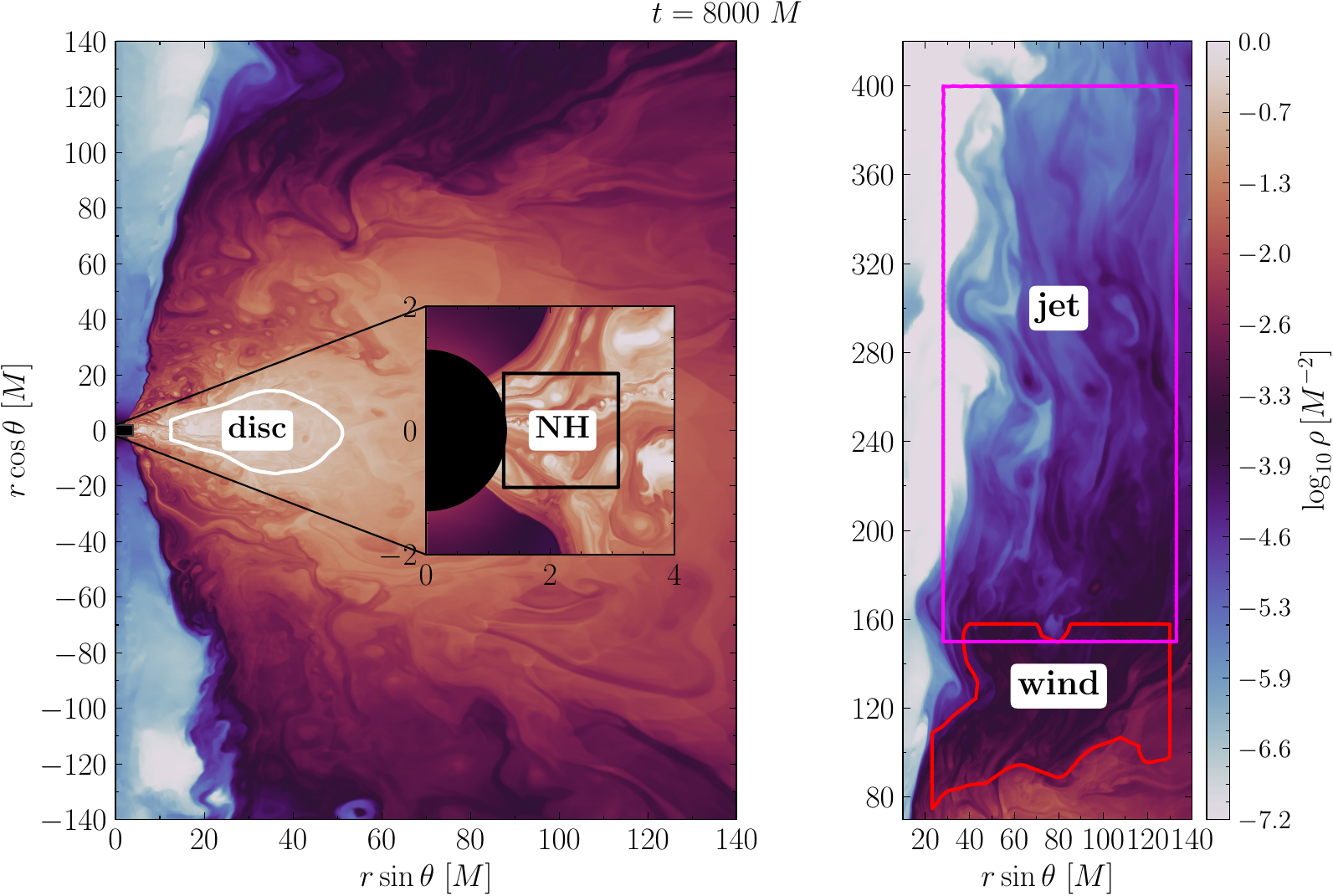}
  \caption{Highlighting of the four different regions in which the
    statistical properties of rest-mass density turbulent field are
    studied. The left panel reports the boundaries of the ``disc'' region
    while the inset zooms-in onto the ``near-horizon'' (NH) region. The
    right panel is the same as on the left but shows the ``wind'' and
    ``jet'' regions, respectively. The data refers to the snapshot at $t
    = 8000 \, M$.}
\label{fig:zones}
\end{figure*}

The evolution of SANE-type accretion torus has been discussed numerous
times in the literature and we will not present it here in detail,
referring the interested reader to the collective presentation by
the~\cite{Porth2019_etal}. On the other hand, because we are here
interested in characterising the turbulent properties of the flow, we
concentrate our attention on the late part of the evolution and analyse
in detail the properties of the accretion flow at five specific times,
namely, $t=8000, \, 8500, \, 9000, \, 9500 \, M$, and $10000 \, M$. At
these times, the dynamics has reached a quasi-stationary evolution and
the turbulent flow is fully developed (see App.~\ref{sec:appendix_A} for
the evolution of the most important quantities characterising the
accretion process).

Figure \ref{fig:fields_8000e10000} offers a view of the evolution by
showing the spatial distribution in a polar slice of some of the most
relevant plasma quantities at two representative times, $t=8000 \, M$
(top row) and $t=10000 \, M$ (bottom row; intermediate times show a very
similar behaviour). More specifically, from left to right, we report the
rest-mass density $\rho$ [panel $(a)$], the temperature $T$ [panel
  $(b)$]\footnote{We compute the temperature as $T = 1.088 \times 10^{13}
( {p}/{\rho} ) \ {\rm K}$, where the numerical factor comes from the
transformation from geometric units to Kelvin~\citep{Zanotti2010,
  CruzOsorio2020}. Note how reconnection in the highly magnetized plasma
($\log_{10} \sigma \geq 2$ ) efficiently heats plasmoids to relativistic
temperatures ($\log_{10} T \geq 12$) \citep{Ripperda2020}.}, the
magnetisation $\sigma: = b^2/\rho$, where $b^2 = b^i b_i$ is the strength
of the magnetic field in the fluid frame [panel $(c)$], and the plasma
beta $\beta := 2p/b^2$, where $p$ is the fluid pressure [panel $(d)$].
Clearly, all fields reported show a turbulent dynamics, with the presence
of vortices at different scales that are either accreted or move to large
distances via a high-magnetisation jet. Importantly, the flow also
exhibits, both on the equatorial plane near the event horizon and along
the sheath of the outgoing jet, the presence of smaller-scale highly
energetic structures, the so-called plasmoids \citep{Fermo2010,
  Uzdensky2010, Huang2012, Loureiro2012, Takamoto2013, Vos2023,
  Imbrogno2024}. These result from the reconnection magnetic-field lines
(\ie the local change of topology of the magnetic field), thus converting
the magnetic energy into internal energy and accelerating particles at
large Lorentz factors. Despite the fact that our simulations are in the
ideal-MHD limit and, numerical resistivity is sufficient to trigger
magnetic reconnection that, together with the generation of plasmoids, is
commonly regarded as a characterising feature of plasma dynamics both at
the microscopic level -- as that explored in particle-in-cell (PIC)
simulations~\citep{Comisso2018, Parfrey2019, MellahEA22, Meringolo2023,
  Vos2024, Imbrogno2025} -- and at the macroscopic level explored by
global MHD simulations~\citep{Servidio2009}.

A more careful inspection of Fig.~\ref{fig:fields_8000e10000} reveals
that the turbulent dynamics is homogeneous but only in specific and
distinct regions that are characterised by the different level of the
rest-mass density, magnetisation and plasma-$\beta$. To reflect these
intrinsic differences, our analysis hereafter will concentrate on four
distinct regions, namely, the ``near-horizon'' (NH), the ``disc'', the
``wind'', and the ``jet''. More specifically, the near-horizon region is
governed by crucial general-relativistic effects and it is where the
influence of the metric on turbulence becomes most significant and cannot
be neglected. The disc, on the other hand, is characterised by a region
of high rest-mass density -- and hence moderate magnetisation and high
plasma beta -- with sustained turbulence over a large dynamical range.
The wind serves as a transitional interface, where the plasma density
reaches average values and where current layers are observed propagating
outwards~\citep{Nathanail2020, Ripperda2020, NathanailEA22}. Finally, the
jet is defined by its lower rest-mass density and a dominant magnetic
field with low plasma beta. In this region, turbulence becomes highly
anisotropic and is suppressed due to strong magnetisation. This
classification in four different zones is crucial to our analysis and all
of our results will be presented in a differential manner for each of
these different regions.

Although the methodology presented in Sec.~\ref{sec:2} to characterise
the turbulent properties is equally applicable to any scalar field
describing the accretion process, hereafter, we will focus our analysis
using as proxy the rest-mass density field and by characterising its
turbulence properties in the four distinct regions mentioned
above. Indeed, the rest-mass density is customarily used for identifying
potential turbulence scaling laws, both in hydrodynamics and in
weakly-compressible plasmas, such as the solar wind~\citep{Frisch1995,
  Bruno2005}.

Given a turbulent rest-mass density field at one of the representative
times mentioned above, and in order to select the different regions, we
first ``smooth'' performing volume averages over spheres of coordinate
radius $\mathcal{R} = 7 \, M$ to obtain a coarse-grained field
$\bar{\rho}_{\mathcal{R}}$. The precise value chosen for $\mathcal{R}$
has little influence on the properties of the smoothed field
$\bar{\rho}_{\mathcal{R}}$ and has been chosen here as a reasonable
compromise between small- and large-scale features in the turbulence.
Next, we determine the spatial boundaries of the four analysis zones
employing the following prescription for the position, or rest-mass
density, or both:
\begin{itemize}

\item ``NH'': \quad $r\sin\theta/M < 10$ and
  $\bar{\rho}_{\mathcal{R}}/M^{-2} > 10^{-1.06}$; 

\item ``disc'': \quad $r\sin\theta/M > 12.5$ and
  $\bar{\rho}_{\mathcal{R}}/M^{-2} > 0.22$;
  
\item ``wind'': \quad $10^{-4} < \bar{\rho}_{\mathcal{R}}/M^{-2} <
  10^{-2.73}$;
  
\item ``jet'': \quad $28 < r\sin \theta /M <133$ and $150 <r\sin
  \theta/M < 400$.

\end{itemize}
Obviously, the criteria for the distinction in four zones is largely
arbitrary but plays little role in our analysis as different
prescriptions would lead to very similar statistical properties (see
discussion in App.~\ref{sec:appendix_B}).

The four different turbulent regions are marked by closed contours in
Fig.~\ref{fig:zones} for the snapshot $t = 8000 \, M$, with regions on
small scales on the left panel (near-horizon and disc) and regions on
large scales on the right panel (wind and jet); an identical procedure is
followed also for the snapshots at different times. As we will show
below, the analysis will reveal that the assumption of homogeneous
turbulence is indeed valid in all of these regions. It is important to
stress that the smoothed rest-mass density field
$\bar{\rho}_{\mathcal{R}}$ is employed only to set the boundaries of the
various zones and that the ensemble averages appearing in
Eq.~\eqref{eqn:S2gamma+geo} are actually done on the unfiltered rest-mass
density field $\rho$.

\section{Comparative measurements of turbulence properties}
\label{sec:pls}

Having introduced in the previous sections the basic principles of the
calculation of the turbulence properties in curved manifolds, we now
discuss some of the more technical aspects that are encountered when
wanting to employ in practice the approaches discussed above. In
particular, when processing the data produced from GRMHD simulations the
challenges to be addressed involve: \textit{(i)} the existence of irregular
boundaries marking the different regions of homogeneous turbulence; \textit{(ii)}
the inherent non-trivial spatial metric characterising any constant-time
slice; and \textit{(iii)} the distribution of grid points where the relevant
quantities are stored and that is irregular either because of the use of
non-trivial coordinate systems (\eg the MKS coordinates) or refinement
levels as those commonly employed in codes such as \texttt{BHAC}.

The calculation of the power spectral density
$\textrm{PSD}_{\mathcal{P}}$ in Eq.~\eqref{eqn:PLS} addresses issues
\textit{(i)} and \textit{(ii)} rather naturally as it relies on the
computation of structure functions and auto-correlation functions, which
are independent of the geometry of the averaging volume, and makes use of
the local value of the three-metric. At the same time, issue
\textit{(iii)} and the complications associated to an irregular
distribution of grid points can be resolved by using a cubic
interpolation over a grid of $1024 \times 1024$ points and based on the
Clough-Tocher method, which employs piecewise polynomial interpolants to
ensure $\mathscr{C}^1$ smoothness and curvature-minimising
properties~\citep[see, e.g.,][]{Alfeld1984, Renka1984}.

\begin{figure}
  \includegraphics[width=0.95\columnwidth]{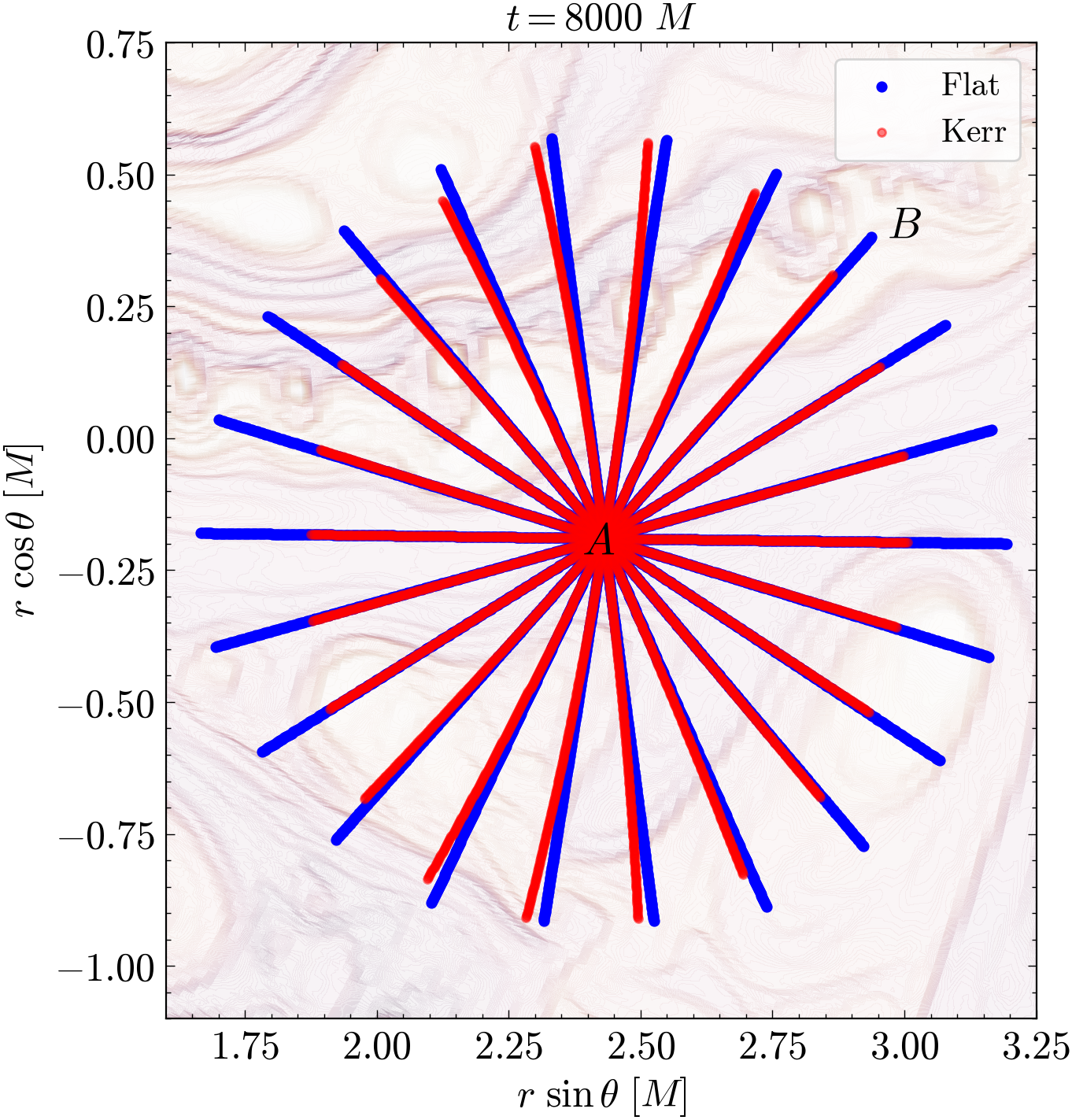}
  \caption{Representative example of the procedure to build the
    second-order structure function starting from a given point $A$ and
    reports with red solid lines the set of curves leading to points $B$
    at a proper distance $\ell=0.76\,M$ from $A$, while the turbulent
    rest-mass density field is shown as a transparent background. To
    create a contrast, we also show with blue solid lines the
    corresponding curves in a flat spacetime. The data refers to the
    snapshot at $t = 8000 \, M$.}
  \label{fig:properlenghtsFK}
\end{figure}

As a first step in computing the $\textrm{PSD}_{\mathcal{P}}$, we need to
assess the set of points at a constant given proper length. We begin by
focusing on the near-horizon region, as highlighted on the left panel of
the Fig.~\ref{fig:zones}, where the curvature effects are most
pronounced. As an illustrative example, and starting from an initial
point $A$, we compute proper lengths directed toward different
surrounding target points $B$, which are separated from $A$ by a finite
distance $\ell$. The calculation of the final positions ends either when
we reach exactly the distance $\ell$ from $A$ to a target point $B$ (this
represents the vast majority of the cases), or when we reach the
boundaries of the mask domain. In this latter case, we resize $\ell$ so
that the domain boundary is not crossed. For each point $A$, this
procedure is repeated by considering $N_\theta = 22$ radial directions at
equally spaced angles for $A$ and we have used $N_A \simeq 10^6$ initial
points, \ie all of $1024 \times 1024$ cells of our interpolated data. We
also tested the sensitivity of our results to the number of angular
directions $N_\theta$ (we recall that $N_\theta = 22$ is the default) and
found consistent results even for $N_\theta = 8$, confirming the
robustness and convergence of our statistical analysis.

Figure~\ref{fig:properlenghtsFK} offers a representative example of our
procedure to build the second-order structure function starting from a
point $A$ located at $(r\sin\theta, r\cos\theta) = (2.4,-0.2) \, M$ and
reports with red solid lines the set of curves leading to points $B$ at a
proper distance $\ell=0.76\,M$ from $A$, while the turbulent rest-mass
density field is shown as a transparent background. Also shown with blue
solid lines are the corresponding curves in a flat spacetime and
comparing the two sets it is easy to appreciate the considerable
differences that emerge when considering proper and coordinate
lengths. Note that the blue curves are ``straight lines'' as they
  represent simple radial-coordinate distances at a constant value of the
  polar coordinate; the red lines, on the other hand, are proper lengths
  between two points having a proper distance $\ell$ and are not
  ``straight lines'' because of the underlying curvature of the spatial
  slice. Next, for each of the regions discussed in
Sec.~\ref{subsec:zones} (see also Fig.~\ref{fig:zones}) we compute the
proper second-order structure function $S_{2, \mathcal{P}}(\ell)$ by
varying the proper length $\ell$ and thus progressively increasing the
integration volume [see Eq.~\eqref{eqn:S2gamma+geo}] until the intended
zone is fully covered.

\begin{figure*}
  \includegraphics[width=0.95\textwidth]{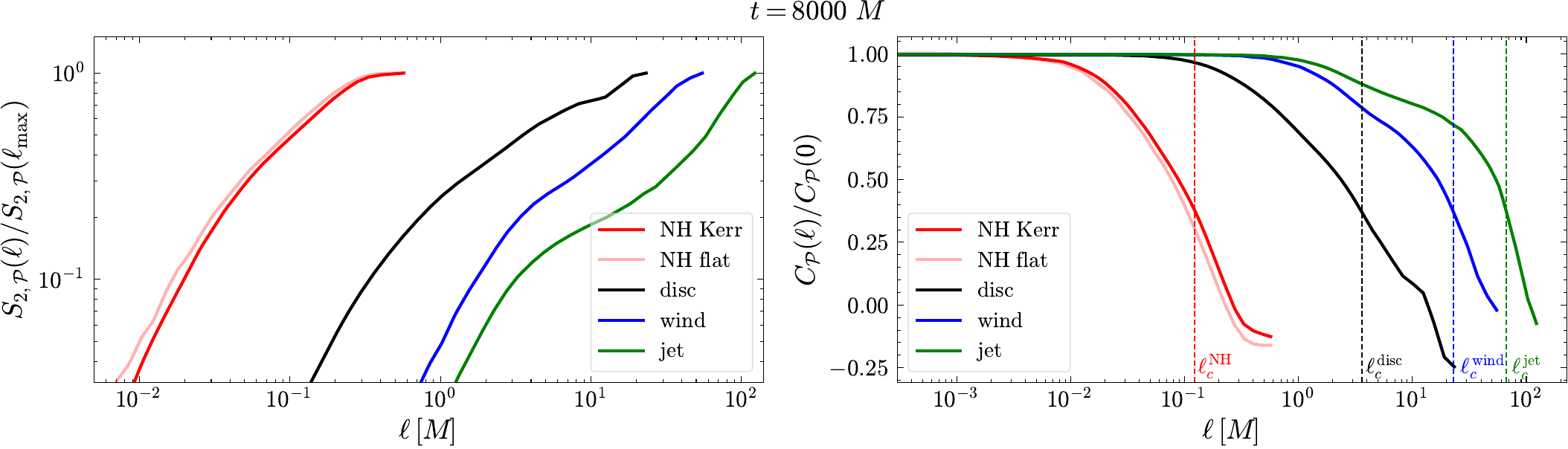}
  \caption{\textit{Left panel:} proper second-order structure functions
    normalised to the maximum proper length measured, \ie $S_{2,
      \mathcal{P}}(\ell)/S_{2, \mathcal{P}}(\ell_{\rm max})$, as computed
    for the NH region (full-red line), the disc (black line), the wind
    (blue line), and the jet region (green line). Also shown as a
    comparison the classical second-order structure function for the NH
    region (light-red line). \textit{Right panel:} using the same
    notation as on the left, we report the proper auto-correlation
    functions normalised by the variance $C_{\mathcal{P}}(\ell) /
    C_{\mathcal{P}}(0)$. The data refers to the snapshot at $t = 8000 \,
    M$.}
    \label{fig:S2_C}
\end{figure*}

The left panel of Fig.~\ref{fig:S2_C} reports the behaviour of the proper
second-order structure function normalised to the maximum proper length
measured, \ie $S_{2, \mathcal{P}}(\ell)/S_{2, \mathcal{P}}(\ell_{\rm
  max})$ at time $t = 8000 \, M$ as a function of the proper length (the
smallest coordinate length that can be resolved by the simulations is $l
\sim 2 \times 10^{-3}\,M$ and is very close to the event horizon; because
turbulent structures need to be resolved over several cells, the smallest
proper length in our analysis is $\ell_{\rm min} \sim 9 \times
10^{-3}\,M$). Reported with different colours is the data for the four
distinct regions: NH (red line), disc (black line), wind (blue line), and
jet (green line). A number of comments are worth making. First, note the
clear distinction in scales of the different regions, such that the
proper second-order structure function in the NH region has $\ell_{\rm
  min} \sim 9 \times 10^{-3} \, M$ and $\ell_{\rm max} \sim 0.6 \, M$,
while in the jet region it has $\ell_{\rm min} \sim 1.4 \, M$ and
$\ell_{c, \rm max} \sim 130 \, M$; in the other regions the range is
smoothly restricted in these ranges. Second, in all four zones, a clear
inertial range is evident, characterized by a power-law scaling of $S_{2,
  \mathcal{P}}(\ell)$, which reflects the self-similar nature of the
turbulent cascade~\citep{Frisch1995}. Third, note that at larger scales
the proper second-order structure function exhibits saturation,
indicating that the energy-containing scales are well-resolved within our
domain~\citep{Frisch1995} and further supporting the assumption of
homogeneity~\citep{Matthaeus1982a}. Fourth, at least for the smallest
scales captured in the NH regions, clear differences appear between the
analysis carried out in the Kerr spacetime (full-red line) and the
equivalent one performed assuming the standard flat-spacetime approach
(light-red line); we will comment further on these differences below.
Finally, it should be remarked that the proper second-order structure
functions shown in Fig.~\ref{fig:S2_C} span almost four orders of
magnitude across the different regions.

The right panel of Fig.~\ref{fig:S2_C} provides a complementary
information by showing the proper auto-correlation function directly [see
  Eq.~\eqref{eq:cs2}] also at time $t = 8000 \, M$ and for the different
regions considered as a function of the proper length. The
auto-correlations are normalised by the variance $C_{\mathcal{P}}(0)$,
which are different for the different regions and reflect the different
level of fluctuations observed. These are larger in the NH region
($C_{\mathcal{P}}(0)=3.4 \times 10^{-2} \, M^{-4}$), and steadily decrease as
one moves from the disc ($C_{\mathcal{P}}(0)=6.5 \times 10^{-3} \,
M^{-4}$), over to the the wind ($C_{\mathcal{P}}(0)=2.5 \times 10^{-7} \,
M^{-4}$), and to the jet region ($C_{\mathcal{P}}(0)=4.2 \times 10^{-9}
\, M^{-4}$); the same approach is followed in both the flat and Kerr
spacetimes. This analysis provides a unique method for determining the
characteristic size of energy-containing turbulent structures, or
``correlation length'' $\ell_c$. More specifically, we estimate this
length as the $e$-folding scale of the auto-correlation
function~\citep{Frisch1995, Matthaeus1982b, Servidio2009}
\begin{equation}
{C_{\mathcal{P}}(\ell_c)} := \frac{1}{e}
     {C_{\mathcal{P}}(0)} \,,
\label{eqn:correlationlenght}
\end{equation}
so that $\ell_c$ effectively provides a measure of the characteristic
(proper) length at which the correlation function has reached about $1/3$
of its variance. Not surprisingly, these characteristic proper lengths
vary significantly across zones, with $\ell_c \simeq 0.1 \, M$ for NH
region, $3.6 \, M$ for the disc, $23.0 \, M$ for the wind and $67.1 \, M$
for the jet.

\begin{figure*}
  \includegraphics[width=0.95\textwidth]{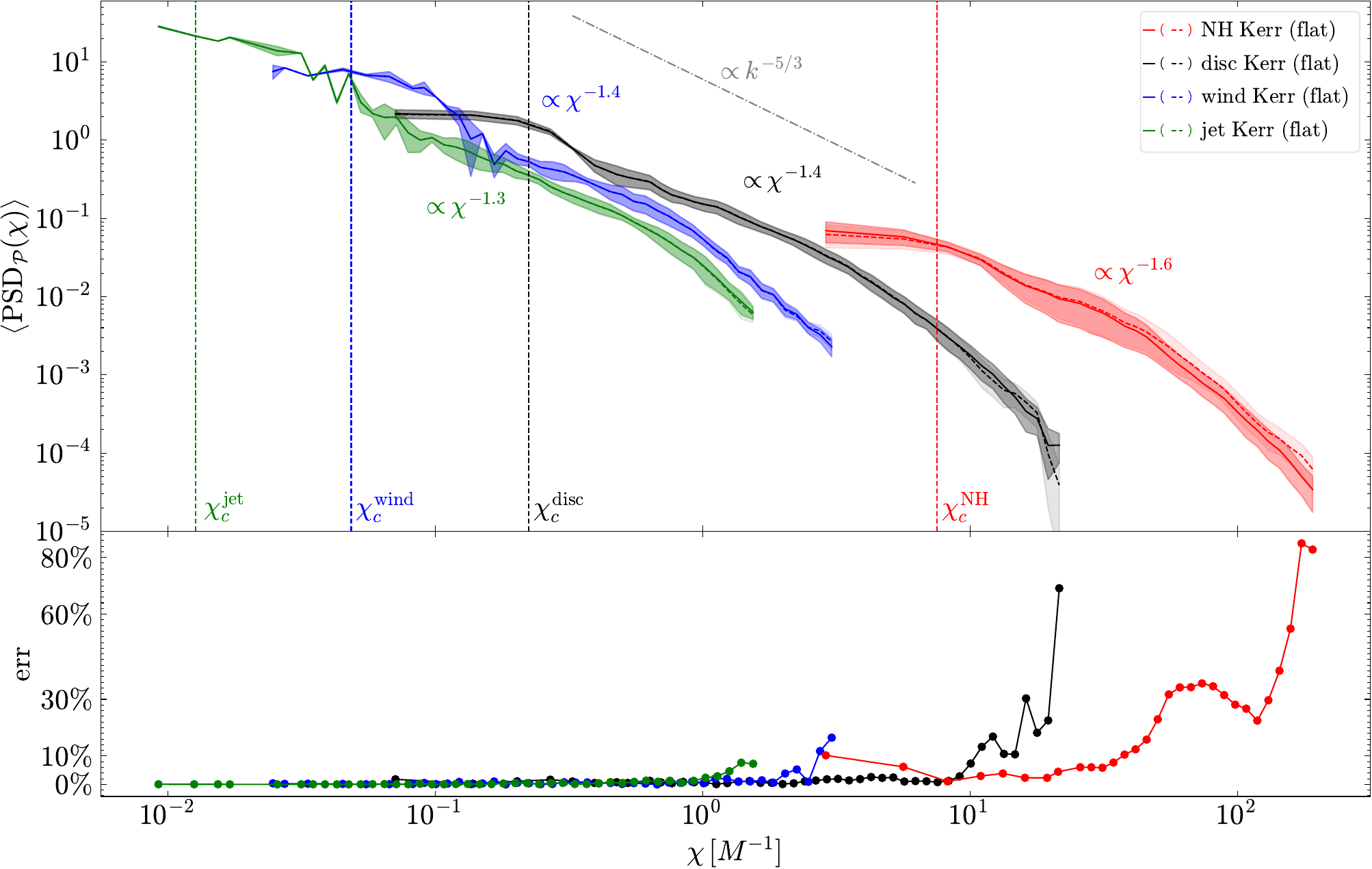}
  \caption{\textit{Top panel:} using the same convention as in
    Fig.~\ref{fig:S2_C} we report the proper power spectral density
    averaged over the five time slices of the numerical simulations
    $\left\langle \textrm{PSD}_{\mathcal{P}}(\chi) \right\rangle$ and
    shown as a function of the wavevector $\chi$; the shaded regions show
    the variability in the PSDs across the five snapshots. Note that, for
    each region, we show the PSDs computed in the case of a curved
    spacetime (solid lines) and for a flat one (dashed
    lines). \textit{Bottom panel}: relative differences in the PSDs
    between a Kerr and a flat metric for each zone. Note that differences
    up to $\sim 40-80\%$ appear in the NH region.}
  \label{fig:PLSerror}
\end{figure*}

At this point, after ensuring the auto-correlation function to be an even
function of the fluctuations, \ie $C_{\mathcal{P}}(\ell) =
C_{\mathcal{P}}(-\ell)$, we can compute the PSD of the auto-correlation
function $\textrm{PSD}_{\mathcal{P}}(\chi, t)$ [see Eq.~\eqref{eqn:PLS}]
for each of the four regions and for each of the five time slices
discussed in Sec.~\ref{subsec:zones} (see Appendix~\ref{sec:appendix_C}
for a validation of the PSD calculation via the use of an analytical
prescription for the rest-mass density); the shaded regions show the
variability in the PSDs across the five snapshots. Furthermore, to ensure
a representative measure of turbulence in the steady state, we average
the spectrum over multiple time intervals $t$. The resulting
time-averaged $\left\langle \textrm{PSD}_{\mathcal{P}}(\chi)
\right\rangle$ are shown in Fig.~\ref{fig:PLSerror}, computed both in the
case of a curved spacetime (solid lines) and for a flat one (dashed
lines).

Not surprisingly, and in analogy with what shown with the second-order
structure function, the PSDs of the different regions span different
ranges in wavenumber, with the jet region covering the largest scales and
with the NH region the smallest ones. Also in this case, the information
collected in this way via the PSDs spans almost four orders of magnitude
across the different regions. Given the definition of the correlation
length $\ell_c$, we can define the correlation wavenumber $\chi_c :=
(\ell_c)^{-1}$ and note that at wavenumbers $\chi \lesssim \chi_c$ the
spectra appear relatively flat. However, a closer inspection of the
$\textrm{PSD}_{\mathcal{P}}$ at wavenumbers $\chi > \chi_c$ reveals the
emergence of a well-defined inertial range, which we identify via the
running slope index
\begin{equation}
\delta_{\text{run}}(\chi) := \frac{d \log \left(
  \textrm{PSD}_{\mathcal{P}}(\chi)\right)}{d \log \chi}\,,
\end{equation}
where $\delta_{\text{run}}(\chi) = \textrm{const.} = -5/3$ corresponds to
the Kolmogorov spectral slope.

Applying this measurement to the jet and wind regions, we find a spectral
scaling of $\textrm{PSD}_{\mathcal{P}} \sim \chi^\delta$, with $\delta
\approx -1.3$ and $-1.4$ respectively, which are flatter than typical
plasma turbulence \citep{Bruno2005}. At smaller scales ($\chi \gg
\chi_c$), an exponential decay is observed and this could either be to
the intrinsic lack of smaller magnetic structure or to the dissipative
mechanisms associated with the GRMHD simulations. The disc region
displays a similar behavior but over smaller scales, where local magnetic
reconnection processes dominate and the spectral slope is the same as the
wind region, $\delta \sim -1.4$.

Finally, the NH region has power on the smallest scales and the PSD
exhibits a clear inertial range with the spectral slope being very close
to a Kolmogorov one, namely, $\textrm{PSD}_{\mathcal{P}} \sim \chi^{-1.6}
\sim \chi^{-5/3}$. This result suggests a striking similarity to
classical turbulence theories at least in the vicinity of the event
horizon but when expressed in terms of proper lengths and proper
volumes. Indeed, the Kolmogorov spectrum is reproduced on a larger
inertial range when the statistical properties are analysed in a curved
spacetime when compared to the equivalent analysis in a flat
spacetime. Arguably, this is among the most important results of our
analysis.

The bottom part of Fig.~\ref{fig:PLSerror} reports with the same colour
coding the relative differences between PSDs computed either in curved or
flat spacetimes, \ie $\text{err} := | 1 -
\textrm{PSD}_{\mathcal{P}}^{\text{flat}} (\chi) /
\textrm{PSD}_{\mathcal{P}}^{\text{Kerr}} (\chi)|$. It is then
straightforward to appreciate that the variance deriving from the two
approaches is rather small in the analysis of the spectra in the jet and
wind regions (green and blue circles) and $\lesssim 15\%$. On the other
hand, rather large dissimilarities appear in the PSDs of the disc and NH
regions (black and red circles), where the differences between the two
spectra can be up to nearly $80 \%$. Overall, the results presented in
the bottom part of Fig.~\ref{fig:PLSerror} underline the importance of a
correct treatment of spacetime-curvature effects when measuring the
properties of homogeneous turbulence in curved spacetimes.

\section{Conclusions}
\label{sec:conclusions}

Turbulence remains a complex and not fully understood phenomenon,
presenting unique challenges already in flat spacetimes and additional
ones when it is considered in strong gravitational fields such as those
in the vicinity of BHs. Obviously, a first step in understanding
turbulence in curved spacetimes is the characterisation of its
statistical properties, which cannot be undertaken employing approaches
that are valid in classical physics and special relativity. To address
this issue, we have introduced a novel analytical framework designed to
study turbulence in curved spacetimes with special attention to the
regions surrounding compact objects like BHs. Our approach is based on
the computation of structure functions adapted to generic manifolds,
making it applicable, in principle, to any curved spacetime.

More specifically, when considering the results of numerical simulations
exploring GRMHD turbulence in a $3+1$ decomposition, we extend the
classical definition of the second-order structure function, and the
associated auto-correlation function -- both of which are commonly
adopted to characterise the properties of turbulence -- by introducing
correlations of the fluctuations across points in the turbulent fields
that are at a constant proper-length separation and not at a constant
coordinate length, as done in classical studies. At the same time, the
volume integrals of the second-order structure function and of the
auto-correlation function are not carried out in terms of coordinate
volumes but in terms of proper volumes. These extensions, that are
incorporated in terms of the spatial three-metric of the associated
time-constant hypersurface on which the numerical data is computed, allow
us to measure the properties of turbulence -- and in particular of the
PSD characterising the distribution of turbulent energy across proper
wavevectors $\textrm{PSD}_{\mathcal{P}}(\chi)$ -- across the whole
numerical domain, starting from the regions near the BH and up to the
weak-field region near the outer boundary.

As a first practical application of our approach, we have explored the
turbulence produced in GRMHD simulations of SANE-accretion models onto a
Kerr BH. Because of the large disparity in scales, we have decomposed our
analysis in four distinct regions, namely, the near-horizon (NH) region,
the disc, the wind, and the jet, as they have distinct turbulence
properties and, more importantly, probe regions with significantly
different curvature. In this way we were able to demonstrate that the new
approach can capture the typical behavior of an inertial-range cascade
that is expected from classical turbulence and that this behaviour is
present both in the high-curvature NH region but also in the
mild-curvature jet region. More importantly, we have found that
differences up to $40-80\%$ emerge in the vicinity of the event horizon
with respect to the standard flat-spacetime approach. While these
differences tend to disappear at larger distances, where differences in
coordinate and proper lengths become smaller, our study highlights that
curvature effects are important in characterising turbulence and that
special care needs to be paid when analysing turbulence in strongly
curved spacetimes.

While this work is meant to lay the ground of a novel approach to measure
the properties of homogeneous turbulence in curved spacetimes, it can be
extended in a number of ways. First, by applying it to numerical
simulations of accretion onto BHs in three spatial dimensions. Second, by
exploring different accretion modes, \eg MAD or alternate poloidal
polarity so as to assess if larger/smaller deviations are present in
those cases. Third, by evaluating the properties of turbulence in
spacetimes that are not as extreme as those near BHs but where turbulence
plays a fundamental role. A good example in this respect are the
turbulent motions encountered in the remnant of a binary neutron-star
merger~\citep{Baiotti2016, Radice2024}. Finally, by applying the
formalism to contexts that are not those of GRMHD simulations but rather
of particle-in-cell simulations, where turbulent motion is starting to be
evaluated~\citep{Meringolo2023, Imbrogno2024} also in curved
spacetime~\citep{Parfrey2019, Vos2024, Meringolo2025a}. While we
  plan to explore these extensions in future works, a more substantial
  progress will be achieved when the issue of turbulence is explored in a
  fully covariant and four-dimensional framework, possibly employing
  four-dimensional Fourier transform in Riemann normal
  coordinates~\cite[see, \eg][]{ParkerToms2009, Calzetta2025,
    Calzetta2025b}.

\section*{Acknowledgements}

We are grateful to the reviewer, Prof. E. Calzetta, for his
  insightful comments and suggestions that have improved the
  presentation.  R. M.  and S.S. acknowledge the supercomputing resources and
support provided by ICSC - Centro Nazionale di Ricerca in High
Performance Computing, Big Data and Quantum Computing - and hosting
entity, funded by European Union-Next Generation
EU. A. C. O. acknowledges DGAPA-UNAM (grant IN110522) and the Ciencia
Básica y de Frontera 2023–2024 program of SECIHTI México (projects
CBF2023-2024-1102 and 257435). G. F. gratefully acknowledges the support
of University of Calabria through a research fellowship funded by DR
1688/2023. M. I. acknowledges the European Union's Horizon Europe
research and innovation programme under grant agreement No. 101082633
(ASAP). Support comes from the ERC Advanced Grant ``JETSET: Launching,
propagation and emission of relativistic jets from binary mergers and
across mass scales'' (Grant No. 884631). L.~R. acknowledges the Walter
Greiner Gesellschaft zur F\"orderung der physikalischen
Grundlagenforschung e.V. through the Carl W. Fueck Laureatus Chair.
S. S. acknowledges the ISCRA Class B project ``KITCOM-HP10BB7U73''.
Authors acknowledge computational support from the Alarico HPPC Computing Facility
at the University of Calabria.

\section*{Data Availability}
Data supporting the findings of this study are available from the authors upon reasonable request.




\bibliographystyle{mnras}
\bibliography{biblio} 




\appendix



\section{Properties of the accretion process}
\label{sec:appendix_A}

\begin{figure}
  \includegraphics[width=0.95\columnwidth]{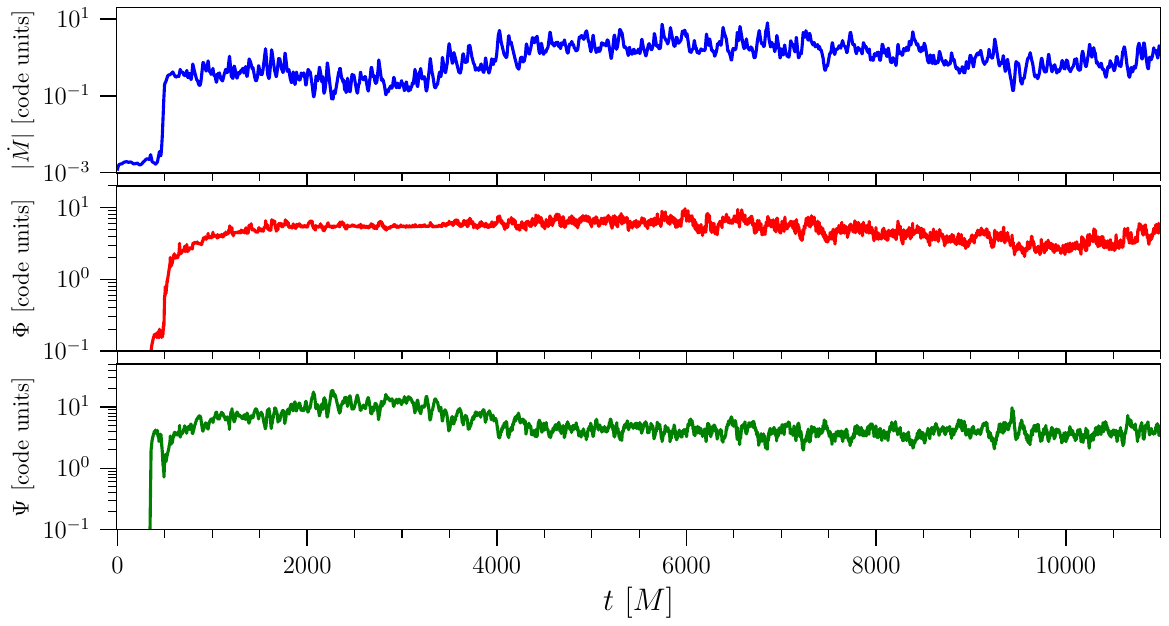}
  \caption{Evolution of the key accretion quantities: the rest-mass
    accretion rate $| \dot{M} |$ (top panel), the accreted magnetic flux
    across the horizon $\Phi$ (middle panel), and the dimensionless MAD
    flux parameter $\Psi := \Phi/\sqrt{\dot{M} }$ (bottom panel).}
  \label{fig:rates}
\end{figure}

During the evolution of the GRMHD simulation, it is a good practice to
monitor key accretion quantities, including the rest-mass accretion rate,
$| \dot{M} |$, the accreted magnetic flux across the horizon, $\Phi$ and,
at the end, the dimensionless MAD flux parameter, $\Psi := \Phi/\sqrt{
  \dot{M} }$. More specifically, the accretion rate and the magnetic flux
threading the horizon are defined as ~\citep{Porth2017}
\begin{eqnarray}
\dot{M} &:=& \int_0^{2\pi} \int_0^{\pi} \rho u^r \sqrt{-g} d \theta d
\phi\,, \\
\Phi &:=& \frac{1}{2} \int_0^{2\pi} \int_0^{\pi} | B^r | \sqrt{-g} d
\theta d \phi\,.
\end{eqnarray}

Figure~\ref{fig:rates} reports the evolution of these quantities up to
$t=10000\,M$ and clearly indicates that the simulation reaches a
quasi-stable accretion rate around $t = 7000 \, M$.  Note also that the
MAD flux parameter remains below the critical threshold value of $\Psi
\simeq 15$~\citep{TchekhovskoyEA2011}, which is consistent with the
classification of the accretion as SANE. 

\section{Impact of the selection of the zone boundaries}
\label{sec:appendix_B}

As discussed in Sec.~\ref{subsec:zones} and illustrated in
Fig.~\ref{fig:zones}, the intrinsically different scales over which
turbulence develops in an accretion process requires the characterisation
of different flow regions (\ie NH, disc, wind and jet), whose boundaries
are necessarily irregular as a response to the defining criteria. Given
this is an important procedural step, it is reasonable to ask whether it
can impact the consequent analysis. To address this question we have
investigated the impact of performing the analysis when the region
studied has boundaries that are either irregular or are set by constant
coordinate lines. Of course, this comparison cannot be exact as
inevitably one choice will include (exclude) parts of the turbulent field
that are absent (present).

\begin{figure}
  \includegraphics[width=0.95\columnwidth]{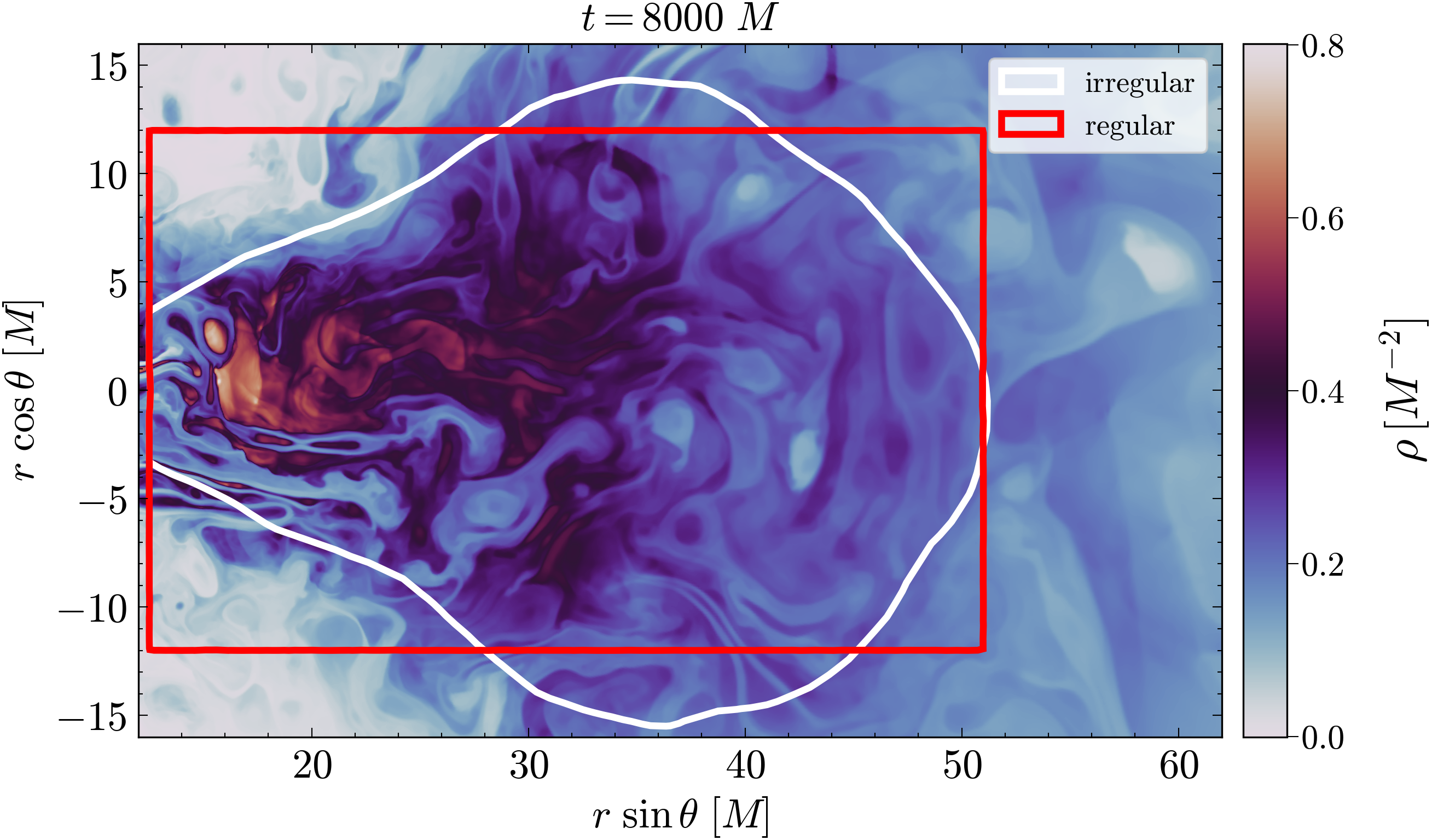}
  \caption{Spatial distributions in a polar slice of rest-mass density
    reported in a linear scale (see also panel $(a)$ of
    Fig.~\ref{fig:fields_8000e10000}) and two regions isolating the disc,
    either with irregular boundaries (white solid line) or with constant
    coordinate lines (red solid line). The data refers to time $t=8000 \,
    M$.}
  \label{fig:diskreg&non}
\end{figure}

Bearing this in mind, and concentrating on a single time $t=8000 \, M$ to
amplify potential differences, we analyse the disc region reported with a
white boundary in Fig. \ref{fig:zones}, and cover it also with a mask
that has regular boundaries given by constant coordinate lines. This is
shown Fig.~\ref{fig:diskreg&non} where the irregular boundary is still
shown in white, while the regular one with a red rectangle. Note that the
overlap is overall good but not perfect and that, in particular, the
regular region encompasses also parts of jet, where the density is
significantly smaller. Fortunately, and as we will comment below, this
difference introduces only an overall difference in the variance, \ie
$\mathcal{C}_{\mathcal{P}}(0)$, and hence a simple scaling in the PSDs.

\begin{figure}
  \includegraphics[width=0.45\textwidth]{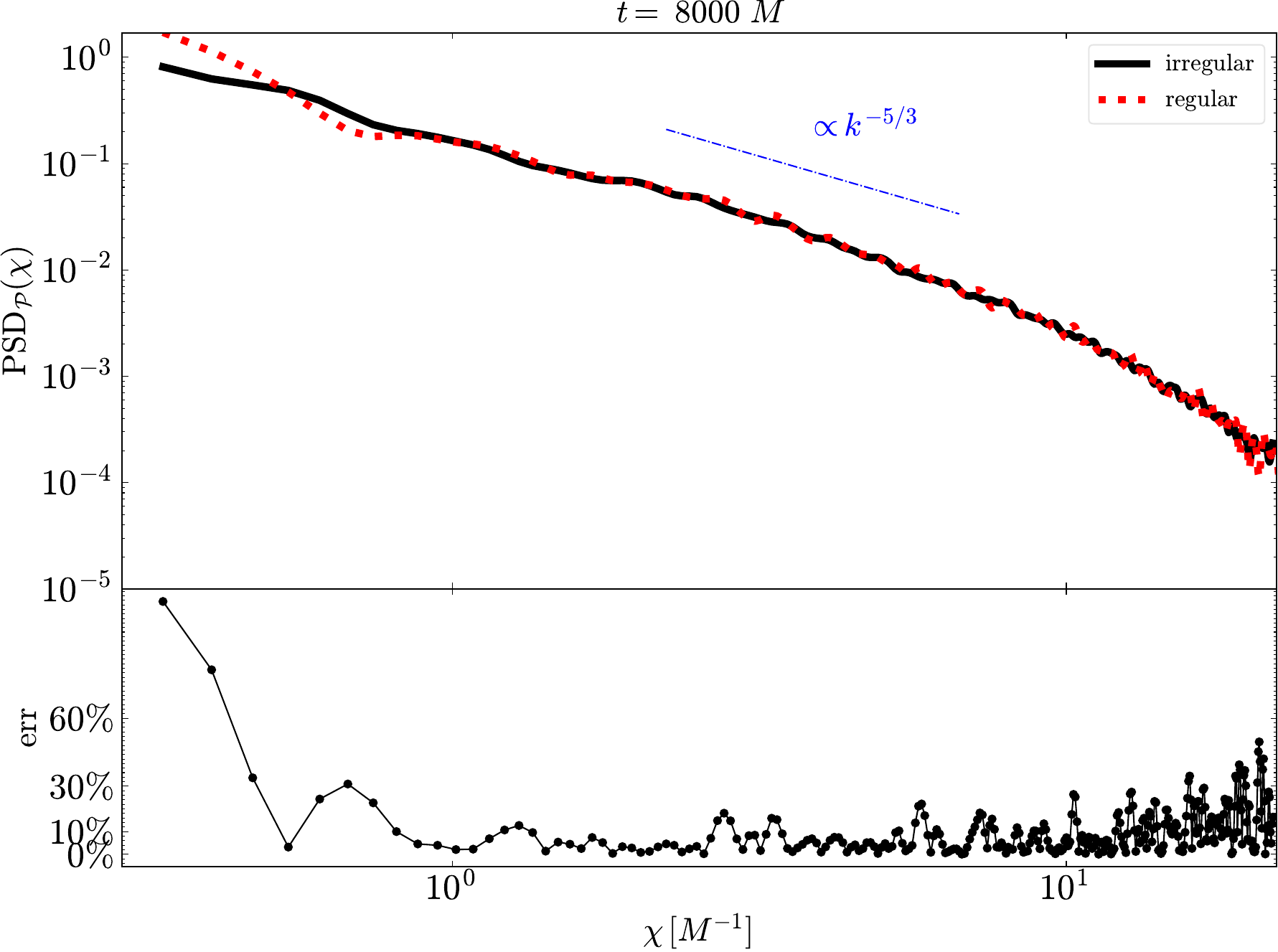}
  \caption{\textit{Top panel:} proper PSDs of the irregular-shape region
    (black solid line) and of the regular-shaped one (red dashed line) at
    a single time $t=8000 \, M$. \textit{Bottom panel:} relative
    differences in the PSDs.}
  \label{fig:DISKspectra}
\end{figure} 

The results of this comparison are reported in the top panel of
Fig.~\ref{fig:DISKspectra}, which shows the proper PSDs of the
irregular-shape region (black solid line) and of the regular-shaped one
(red dashed line) after a rescaling of a factor $\sim 3/2$ due to the
sampling of large-scale and low rest-mass densities (\ie for $\chi < 10
\, M^{-1}$). Clearly, the PSDs are very similar and show the same
inertial behaviour, as also quantified in the bottom panel, which reports
the relative differences in the PSDs, \ie $\text{err} := | 1
-\textrm{PSD}_{\mathcal{P}}^{\text{irreg}} (\chi) /
\textrm{PSD}_{\mathcal{P}}^{\text{reg}} (\chi)|$. Overall, the results in
Fig.~\ref{fig:DISKspectra} confirm the expectation that the choice of
boundaries in the various regions does not affect the turbulence
properties apart from a scaling factor that is ignored in our analysis
(see Fig.~\ref{fig:PLSerror}).

\section{Validation of the PSD extraction}
\label{sec:appendix_C}

\begin{figure*}
  \includegraphics[height=0.42\textwidth]{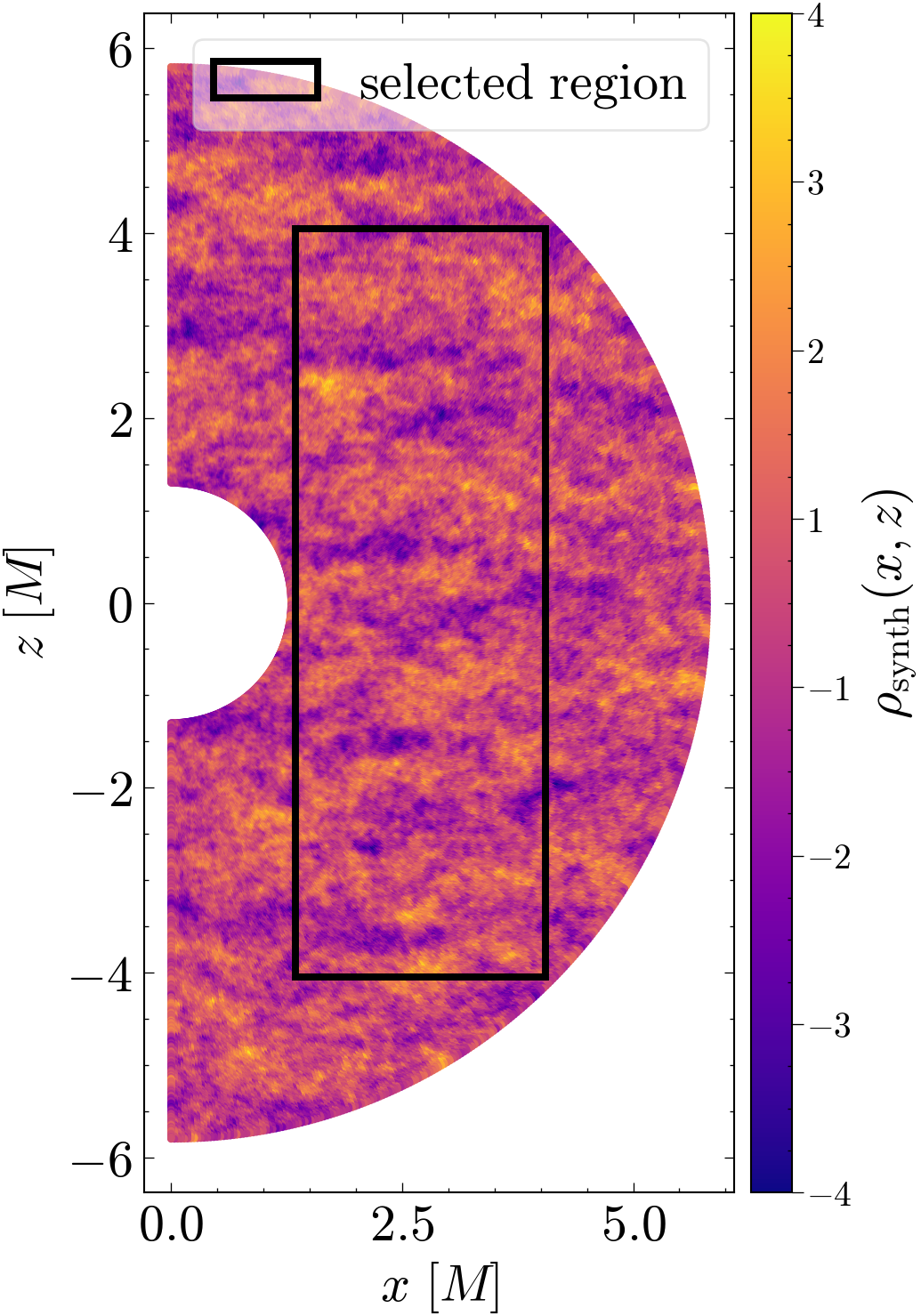}
  \hskip 0.5cm
  \includegraphics[width=0.55\textwidth]{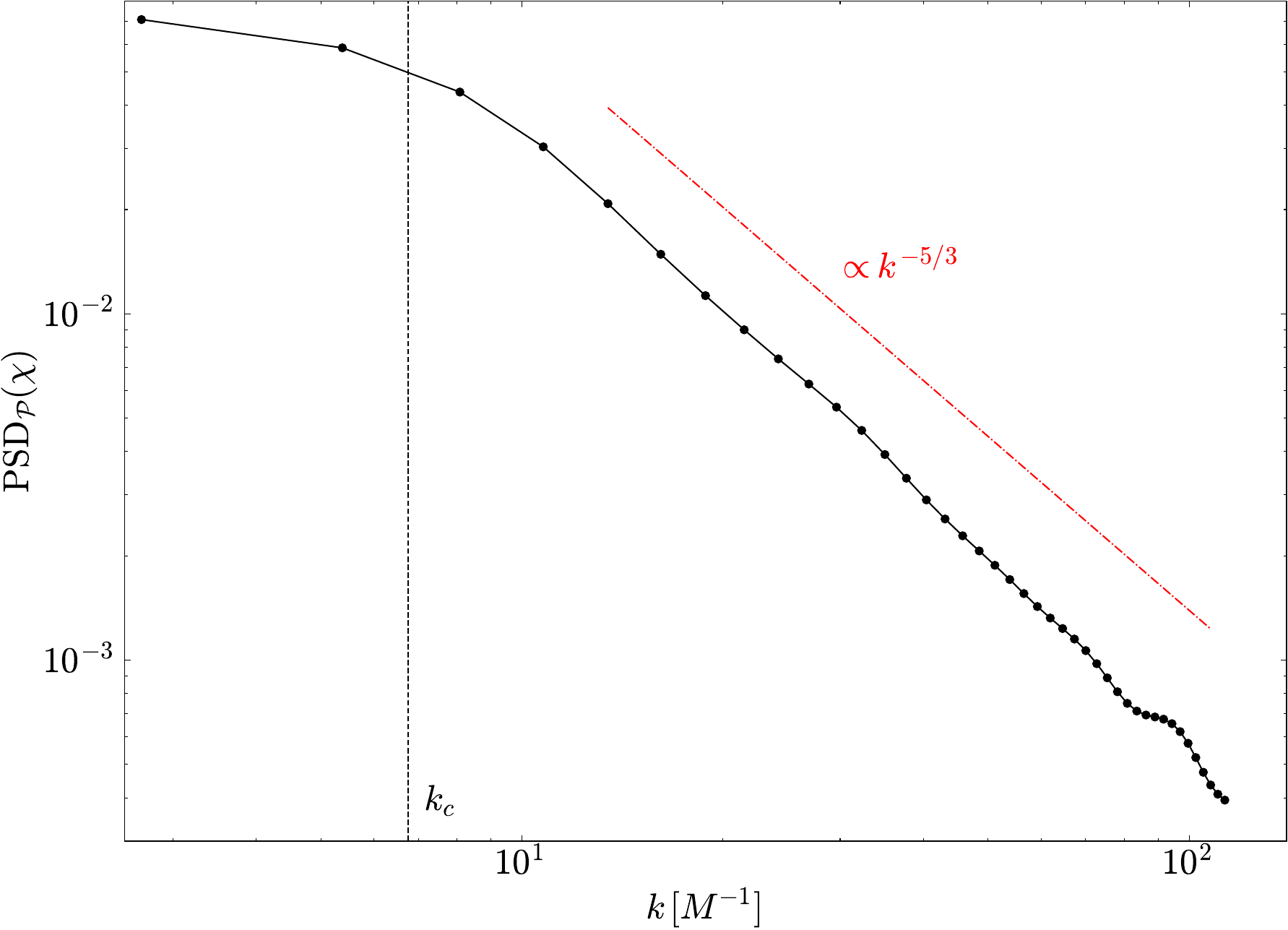}
  \caption{\textit{Left panel:} Spatial distribution of the synthetic
    rest-mass density profile $\rho_{\rm synth}(x, z)$ and the
    rectangular region selected to perform the statistical analysis
    (black solid line). \textit{Right panel:} The PSD obtained from the
    analysis pipeline confirming the expected $k^{-5/3}$ power law. The
    dashed vertical line indicates the correlation wavevector $k_c$ with
    the corresponding correlation length that is $l_{c}=0.15 \, M$.}
  \label{fig:kolmofield}
\end{figure*}
\begin{figure*}
  \begin{minipage}{0.49\textwidth}
  \centering
  \includegraphics[width=0.95\textwidth]{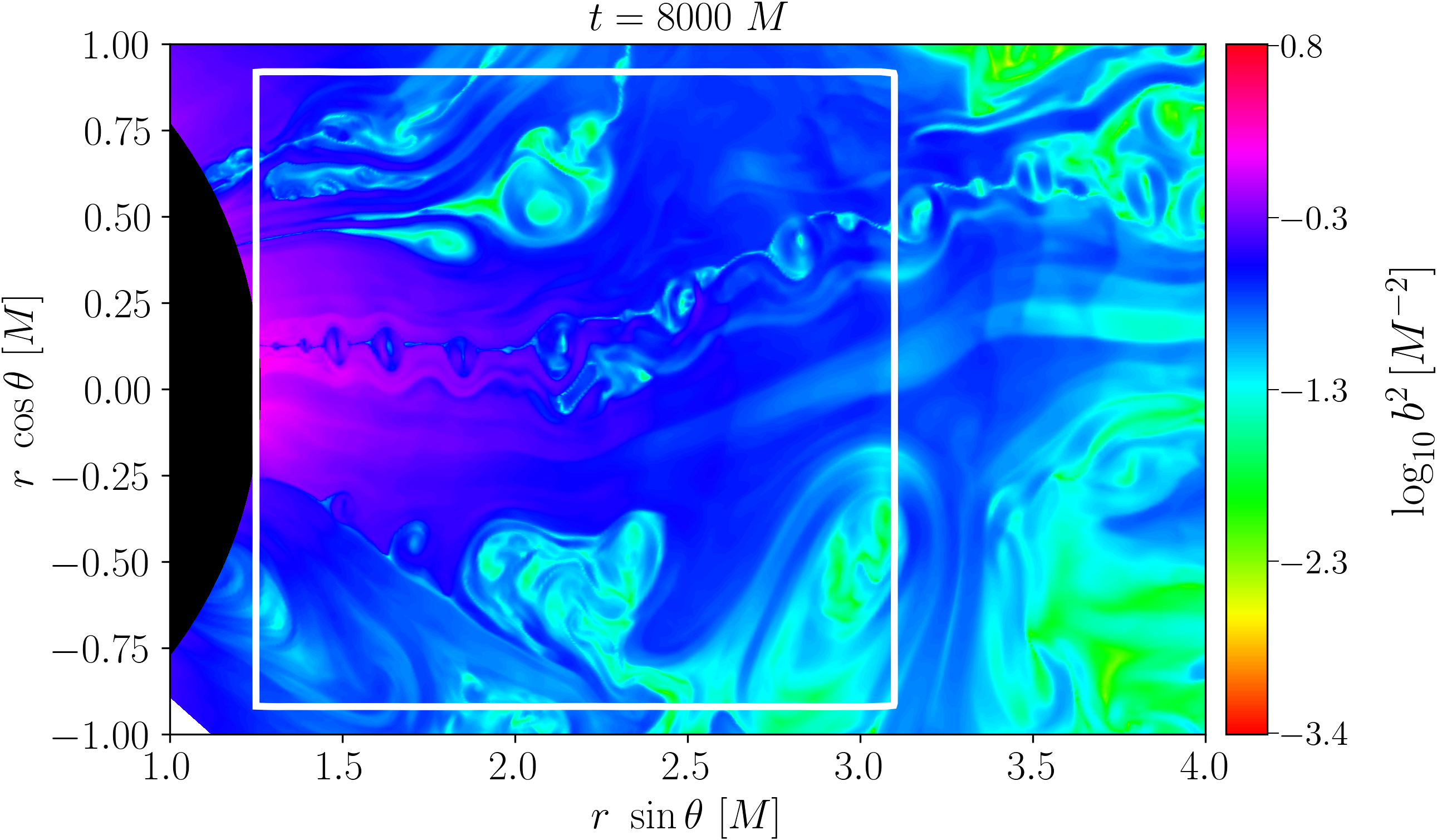}
  \end{minipage}
  \begin{minipage}{0.49\textwidth}
  \centering
  \includegraphics[width=0.8\textwidth]{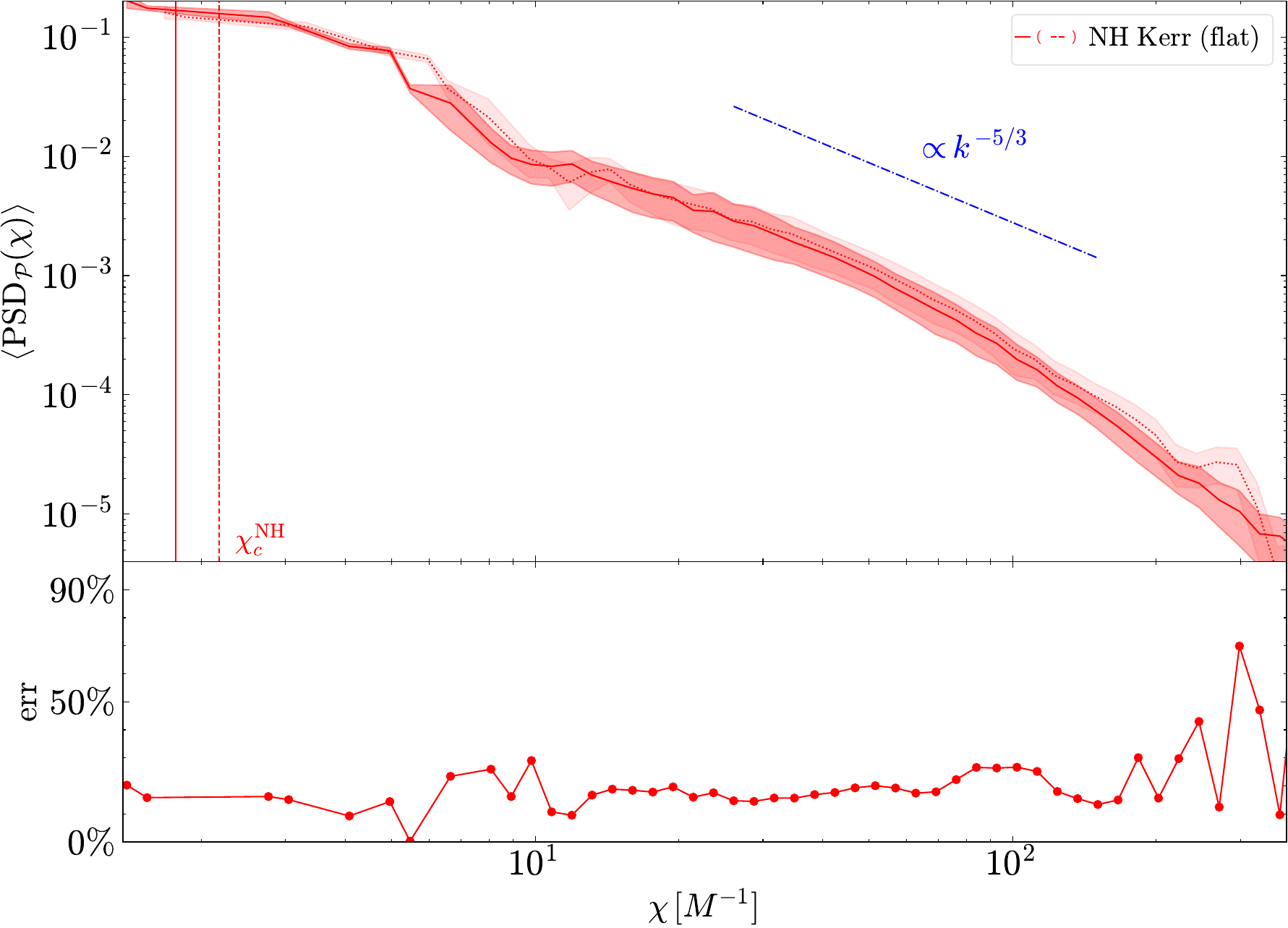}
  \end{minipage}
  \caption{\textit{Left panel:} Spatial distribution in a polar slice of
    the modulus of the comoving magnetic field, with the square selecting
    the NH region where the statistical analysis is performed (white
    solid line). \textit{Right top panel:} Proper PSD averaged over the
    five time slices of the numerical simulations $\left\langle
    \textrm{PSD}_{\mathcal{P}}(\chi) \right\rangle$ in the case of a
    curved spacetime (solid line) and for a flat one (dashed
    line). \textit{Right bottom panel:} relative difference in the PSDs
    between a Kerr and a flat background.}
  \label{fig:modb}
\end{figure*}

As a validation of the procedure outlined in Sec.~\ref{sec:2} for the
calculation of the proper PSD we apply our analysis on a synthetic
turbulent rest-mass density field representing a two-dimensional
Kolmogorov spectrum in flat space with a $k^{-5/3}$ power law and given
by~\citep{Oughton1994} 
\begin{equation}  
  \rho_{\rm synth}(x, z) := \sum_{k_x} \sum_{k_{z}} A_k \, \exp \ [i(k_{x}
    x + k_{z} z + \phi_k)]\,,
\label{eqa}
\end{equation}
where $x := r \sin \theta$ and $z := r \cos \theta$ are the coordinates
in the representative plane, $\bm{\vec{k}}=(k_x,k_z)$ denotes the
wavevector in Fourier space, $\phi_k$ represents randomly selected
phases, and $A_k$ is the amplitude, which we write as~\citep{Oughton1994}
\begin{equation}
A_k := \frac{ k^{-1/2} \, \dfrac{k}{k_0}}{\sqrt{1 + \left(\dfrac{k}{k_0}
    \right)^{11/3}}} \exp\left(-\nu
\left|\left(\frac{k}{{k_{\text{max}}}}\right)^{\nu}\right|\right)\,.
\label{eq:A_k}
\end{equation}
Note that $P(k)$ is proportional to $k |A_k |^2$ and that in the
intermediate range of $k \sim k_0 \ll k_{\text{max}}$, so that $A_k \sim
k^{-4/3}$ and $P(k) \propto k |A_k |^2 \sim k^{-5/3}$, which is the
expected Kolmogorov law.

In expression~\eqref{eq:A_k}, $\nu$ is a cutoff exponent governing the
spectrum for $k \gg k_{\rm max}$, while $k_0$ is the wavevector where the
power spectrum has its maximum. In practice, we have used $\nu=16$,
$k_{\rm max}=10^{3}\, \Delta k$, and $k_0 = 3\, \Delta k$ with $\Delta k
= 2 \pi/L$ and $L= 2\pi M/3$. Periodic boundary conditions are imposed,
restricting the Fourier series to a finite set of modes $k_x$ and $k_z$
and we note that $k_0$ can be seen as linked to the correlation
wavenumber associated with the large-scale correlation length
$k_c$. Indeed, with our choice of parameters $k_0 = 9/M$ and $k_c \simeq
6.7/M$.

The synthetic turbulent field is shown with a colormap in the left panel
of in Fig.~\ref{fig:kolmofield}, where the rectangle with boundaries
$1.35 \, M \leq x \leq 4.05 \, M$ and $|z| \leq 4.05 \, M$ mark the
region (black solid line) over which the analysis is carried out. The
right panel reports with black circles the results of the PSD, which
reproduces well the expected $k^{-5/3}$ power law and thus provides
evidence of the correctness of our analysis pipeline. 

\section{Robustness of the deviations}
\label{sec:appendix_D}

So far, our analysis has concentrated on the statistical properties in
the rest-mass density field and we have highlighted how the use of a
proper measurement is most important in the NH region, where differences
of $40-80\%$ are possible. It is natural to ask whether such deviations
are present also for other scalar quantities, which we expect to be in a
turbulent state, although not necessarily following a Kolmogorov
spectrum. To address this point, we considered the turbulence properties
of the modulus of the magnetic field as measured in the fluid frame,
namely,
\begin{equation}  
  b^2 := \sqrt{\gamma_{ij}b^i b^j}\,.
\end{equation}
Since this is another scalar quantity, no modification is needed to the
analysis presented in the main text, while a more complex analysis --
involving a suitably defined parallel transport -- would be necessary for
a vector field.

The results of this validation are presented in Fig.~\ref{fig:modb},
whose left panel reports the spatial distribution in a polar slice of the
magnetic-field strength $b^2$, with the square selecting the NH region
where the statistical analysis is performed (white solid line). Note
that the field is in this case less turbulent on the largest scales but
also the appearance of an equatorial current sheet and the presence of
plasmoids, that are characteristic of these simulations. The right top
panel, on the other hand, reports the proper PSD averaged over the five
time slices of the numerical simulations $\left\langle
\textrm{PSD}_{\mathcal{P}}(\chi) \right\rangle$ in the case of a curved
spacetime (solid line) and for a flat one (dashed line). This figure,
which should be compared with the equivalent Fig.~\ref{fig:PLSerror} in
the main text, provides the evidence needed, namely, that deviations in
measurement of the turbulent properties in curved and flat spacetimes
emerge independently of the quantity considered. In addition, as reported
in the right bottom panel, the relative difference in the PSDs is again
showing a variation of $\sim 60\%$ at the smallest scales between the two
approaches.


\bsp	
\label{lastpage}
\end{document}